\documentclass[fleqn,usenatbib]{mnras}

\usepackage[T1]{fontenc}
\DeclareRobustCommand{\VAN}[3]{#2}
\let\VANthebibliography\thebibliography
\def\thebibliography{\DeclareRobustCommand{\VAN}[3]{##3}\VANthebibliography}
\usepackage{graphicx,xcolor}	
\usepackage{amsmath}	
\usepackage{booktabs}
\usepackage{amssymb}	
\usepackage{mathtools}
\usepackage{nicefrac} 
\usepackage{newtxtext,newtxmath}
\usepackage{multirow}

\newcommand{\cubepm}{\textsc{CUBEP$^3$M}}
\newcommand{\ccray}{\textsc{C$^2$Ray}}
\newcommand{\nbody}{\textsc{N-body}}
\newcommand{\hmpc}{$h^{-1}$Mpc}

\newcommand{\sphere}{{\sc Sphere}}
\newcommand{\coneone}{{\sc Cone-1}}
\newcommand{\conetwo}{{\sc Cone-2}}
\newcommand{\conethree}{{\sc Cone-3}}

\def\HII{\ion{H}{II}~}

\usepackage{tikz,xcolor,hyperref}
\definecolor{lime}{HTML}{A6CE39}
\DeclareRobustCommand{\orcidicon}{%
	\begin{tikzpicture}
	\draw[lime, fill=lime] (0,0) 
	circle [radius=0.16] 
	node[white] {{\fontfamily{qag}\selectfont \tiny ID}};
	\draw[white, fill=white] (-0.0625,0.095) 
	circle [radius=0.007];
	\end{tikzpicture}
	\hspace{-2mm}
}

\foreach \x in {A, ..., Z}{%
	\expandafter\xdef\csname orcid\x\endcsname{\noexpand\href{https://orcid.org/\csname orcidauthor\x\endcsname}{\noexpand\orcidicon}}
}

\title[Anisotropic Source Emission]{Impact of
anisotropic photon emission from sources during the epoch of reionisation}

\author[Timo P. Schwandt et al.]{
Timo P. Schwandt$^{1}$\thanks{E-mail: timo\_schwandt@hotmail.com},
Ivelin Georgiev$^{2,3}$\thanks{E-mail: ivelin.georgiev@astro.su.se}\orcidA,
Sambit K. Giri$^{4,5}$\orcidB,
Garrelt Mellema$^{3}$\orcidC, and
Ilian T. Iliev$^{1}$\orcidD
\\
$^{1}$Astronomy Centre, Department of Physics \& Astronomy, Pevensey III Building, University of Sussex, Falmer, Brighton, BN1 9QH, United Kingdom\\
$^{2}$ ARCO (Astrophysics Research Center), Department of Natural Sciences, The Open University of Israel, 1 University Road, PO Box 808, Ra’anana 4353701, Israel \\
$^{3}$Department of Astronomy and Oskar Klein Centre, AlbaNova, Stockholm University, SE-10691 Stockholm, Sweden\\
$^{4}$Nordita, KTH Royal Institute of Technology and Stockholm University, Hannes Alfvéns väg 12, SE-106 91 Stockholm, Sweden\\
$^5$Van Swinderen Institute for Particle Physics and Gravity, University of Groningen, Nijenborgh 4, 9747 AG Groningen, The Netherlands
}

\date{Submitted to MNRAS. Received YYY; in original form ZZZ, NORDITA-2025-002}
\pubyear{2025}

\begin{document}
\label{firstpage}
\pagerange{\pageref{firstpage}--\pageref{lastpage}}
\maketitle
\begin{abstract}
The reionisation of the intergalactic medium (IGM) was driven by the first stars, galaxies, and accreting black holes. However, the relative importance of these sources and the efficiency by which ionising photons escape into the IGM remain poorly understood. Most reionisation modelling frameworks assume idealised, isotropic emissions. We investigate this assumption by examining a suite of simulations incorporating directed, anisotropic photon emissions. We find that such anisotropic emissions of ionising photons yield a different reionisation geometry compared to the standard, isotropic, case. During the early stages of reionisation (when less than 30 per cent of the Universe is ionised), simulations with narrow photon leakage channels produce smaller ionised bubbles on average. However, these bubbles grow to similar sizes during the middle stages of reionisation. This anisotropy not only produces a distinctive evolution of the size distribution of the ionised regions, but also imprints a feature onto the spherically averaged power spectra of the 21-cm signal throughout reionisation. We observe a suppression in power by about 10-40 per cent at scales corresponding to wavenumbers $k = 0.1-1 \, h \, \mathrm{Mpc}^{-1}$, corresponding to the range in which current radio interferometers are most likely to measure the power spectrum. The simulation with the narrowest channel of ionisation emission shows the strongest suppression. However, this anisotropic emission process does not introduce any measurable anisotropy in the 21-cm signal.

\end{abstract}


\section{Introduction}

The Epoch of Reionisation (EoR) is the period in the history of our Universe when the first stars and galaxies released ionising photons into the intergalactic medium (IGM), causing a transition from a cold, neutral state into the hot, ionised one in which it has remained ever since 
\citep[see e.g.][for reviews]{pritchard201221,barkana2016rise,mcquinn2016evolution,dayal2018early}. 
Several telescopes have observed galaxies from this era, especially the James Webb Space Telescope (JWST) \citep[e.g.,][]{naidu2022two,robertson2023identification,tang2023jwst,cameron2023jades,2023ApJ...949L..42M,harikane2023comprehensive,donnan2024jwst}. 
This data has helped us constrain structure formation during these early times \citep[e.g.,][]{mason2023brightest,cueto2024astraeus,dayal2024warm,yung2024characterizing}. 

At the same time, radio telescopes are probing for the 21-cm signal produced by neutral hydrogen in the IGM. 
We can learn about the astrophysics \citep[e.g.,][]{greig201521cmmc,dixon2016large,giri2018bubble,schaeffer2023beorn} and cosmology \citep[e.g.,][]{sitwell2014imprint,giri2022imprints,schneider2023cosmological} of the early Universe with this signal. 
Currently, statistical detection of this signal is the key science goal of the Murchison Widefield Array (MWA)\footnote{\url{http://www.mwatelescope.org/}} \citep[e.g.][]{trott2020deep}, the Low-frequency array (LOFAR)\footnote{\url{http://www.lofar.org/}} \citep[e.g.,][]{Mertens2020,Mertens2025lofar,ghara2020constraining}, and the Hydrogen Epoch of Reionization Array (HERA)\footnote{\url{https://reionization.org/}} \citep[e.g.,][]{HERA2021,TheHERACollaboration_2022}.
In the near future, the Square Kilometre Array (SKA) telescope will begin its observations and its low-frequency part will be sensitive enough to map the spatial distribution of the 21-cm signal during EoR \citep[e.g.,][]{mellema2013reionization,giri2018optimal}. Therefore, modelling this signal accurately is crucial to interpreting these datasets.

One of the key ingredients in understanding the reionisation process is the escape of ionising (UV) radiation from the luminous source, as this determines the evolution of the IGM properties. 
While there exist various estimates of the UV escape fraction from low redshift ($z<3$) observations \citep[e.g.,][]{bergvall2006first,leitet2013escape,izotov2018low,izotov2021lyman}, this quantity is still barely constrained for high redshifts \citep[e.g.,][]{mitra2015cosmic,lin2024quantifying,napolitano2024peering,saxena2023jades,saxena2024jades}. 
These constraints are derived by studying features in the spectra of the galaxies \citep[e.g.,][]{zackrisson2011spectral,zackrisson2017spectral,giri2020identifying,begley2024connecting}. 
As these observations cannot resolve the full emission structure around the observed galaxies, such estimates may be biased if the emissions of the photons are found to be anisotropic. 

Several processes, such as scattering and absorption of photons by the interstellar medium within the galaxies, determine the final emission experienced by the IGM \citep[e.g.,][]{razoumov2006escape,gnedin2008escape,trebitsch2017fluctuating}. High-resolution simulations of early galaxies have found such anisotropic leakage of ionising radiation \citep[see ionisation channels in e.g.,][]{cen2015quantifying,paardekooper2015first,trebitsch2017fluctuating}. 
These simulations have mostly focused on the mass-dependence of the cumulative escape of photons. A notable study of the ionising emission anisotropy by \citet{Yeh2023} of the \textsc{THESAN} reionisation simulations \citep{Garaldi2022,Kannan2022,Smith2022}, shows that the escape fraction of the majority of halos of $M_{\mathrm{halo}} < 10^{10}$ M$_{\odot}$ have a large sight-line to sight-line variability across the EoR (see sec.~5.2). However, due to sheer sizes needed to accommodate large-scale variances \citep[e.g.,][]{Iliev2014largebox,kaur2020minimum,giri2023suppressing}, large-scale cosmological reionisation simulations cannot resolve the internal structure of ionising galaxies and, therefore, require assumptions on the escape of ionising photons from galaxies \citep[see e.g.,][]{dixon2016large,park2019inferring,schneider2023cosmological,giri202421}. 
To date, all large-scale ($\gtrsim 100$ Mpc) reionisation simulations have assumed spherically isotropic emission. However, this assumption has not been fully justified or tested. 

The isotropic emission is also an inherent assumption for fast approximate simulation frameworks, such as \textsc{HMreio} \citep[][]{schneider2021halo,schneider2023cosmological,giri2022imprints}, \textsc{21cmFAST}\footnote{\url{https://github.com/21cmfast/21cmFAST}} \citep{mesinger201121cmfast,Murray2020py21cmfast}, \textsc{simfast21}\footnote{\url{https://github.com/mariogrs/Simfast21}} \citep{santos2010fast}, \textsc{21cmSPACE} \citep[e.g.][]{2024MNRAS.529..519G}, \textsc{SCRIPT} \citep{choudhury2018photon}, \textsc{BEARS} \citep{thomas2009fast}, \textsc{GRIZZLY} \citep{ghara2018prediction}, and \textsc{BEoRN}\footnote{\url{https://github.com/cosmic-reionization/BEoRN}} \citep{schaeffer2023beorn}\footnote{Note that the anisotropy in some of these semi-numerical frameworks is broken in the way they account for the overlapping ionised bubbles \citep[e.g.,][]{ghara2018prediction,choudhury2018photon,schaeffer2023beorn}.}. 
As these frameworks were developed to interpret observations, it is important to understand the implications of anisotropic escape of ionising photons on the observables, such as the 21-cm power spectra and Lyman-$\alpha$ forest data. Such a study can only be performed with a full radiative transfer code.

The aim of this paper is to
perform a first exploration of the effect of anisotropic escape of ionising photons on the
large-scale reionisation process. 
Our paper is structured as follows. First, in Sec.~\ref{sec:method} we describe our methodology for modelling the anisotropic leakage of ionising radiation, followed by our results in Sec.~\ref{sec:results} and the summary and conclusions in Sec.~\ref{sec:conclusion}. 

\section{Methodology} \label{sec:method}

In this section, we describe the methodology employed to produce our cosmological reionisation simulations. In Sec.~\ref{sec:method_sources}, we present our source modelling. In large ($\gtrsim$200 \hmpc{}) scale simulations, resolving individual sources is computationally expensive. Therefore we have developed a parametric form to control the direction of photon emission at ($\sim$1 \hmpc{}) scales resolved by our simulations. We present the modelling of isotropic and anisotropic photon escape from the sources in Sec.~\ref{sec:method_spherical} and~\ref{sec:method_conical}, respectively.

\begin{figure}
	\includegraphics[width=\columnwidth]{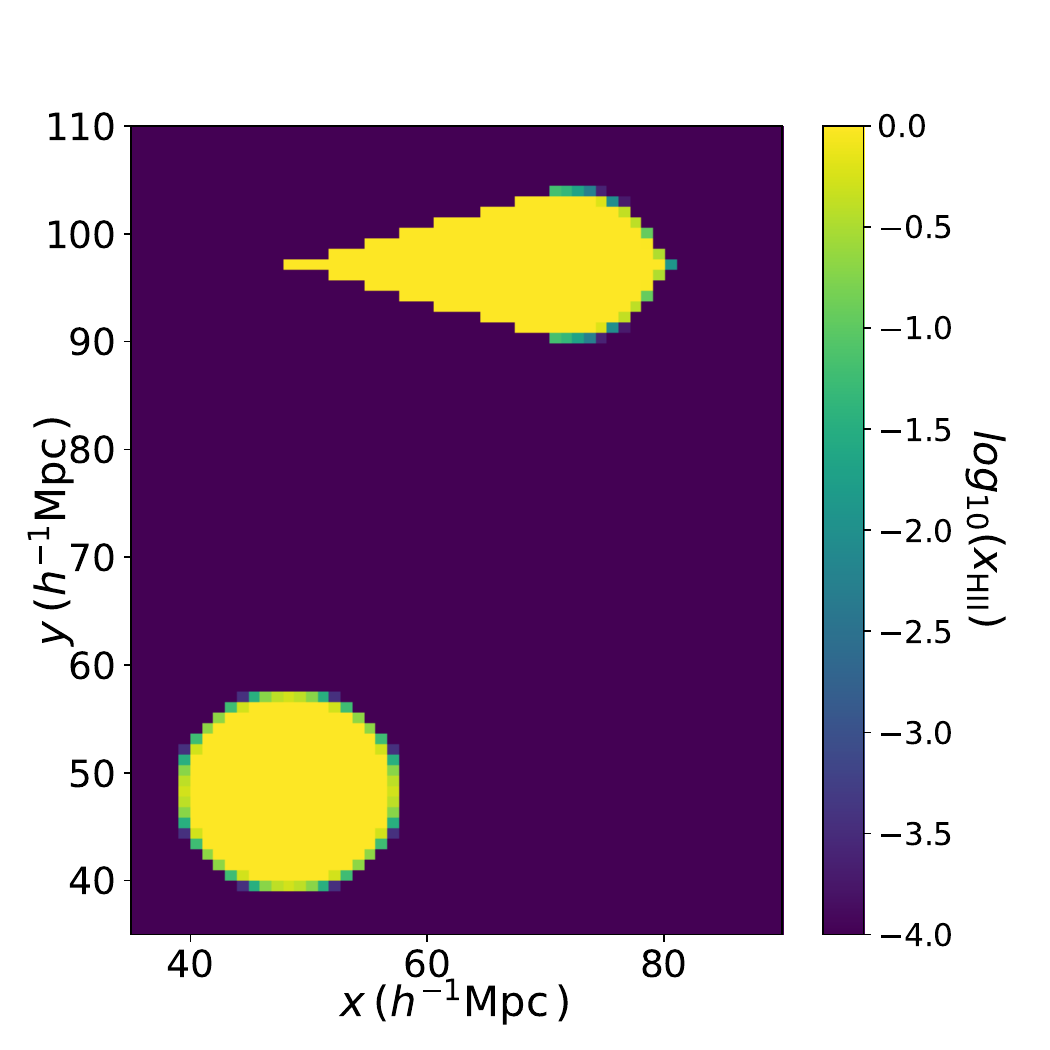}
    \vspace{-2em}
    \caption{
    An illustration of the ionised regions created by two sources with the same photon production rate but different directionality of the photon leakage. The source at the bottom left corner emits photons isotropically creating a spherical ionised region. For the source at the top left corner the emission is restricted to a channel along the x-axis, instead creating a conically-shaped ionised region.
    The panel shows a slice defined by the two sources and the axis of the emission cone. 
    } 
    \label{fig:One_Cone}
\end{figure}

\subsection{Reionisation sources}\label{sec:method_sources}

We simulated cosmological structure formation during the EoR using the \nbody{} code \cubepm{}\footnote{\url{https://github.com/jharno/cubep3m}} \citep{harnois2013}.
We used $4000^3$ particles in a volume of 244 \hmpc{} along each direction. We adopted a flat $\Lambda$-Cold Dark Matter cosmology with parameters
($\Omega_{\mathrm{m}},\Omega_{\mathrm{b}},h,n_{s},\sigma_{8}$) = (0.27, 0.044, 0.7, 0.96, 0.8), where and 
$\Omega_{\mathrm{b}}$ and $\Omega_{\mathrm{m}}$ are the density parameters for baryonic and dark matter, compatible with the  \textit{WMAP} \citep{hinshaw13} and \textit{Planck} results \citep{planck14}. 
Simulation snapshots were saved at every 11.5 Myrs where matter distribution was smoothed to $250^3$ grids. 
The dark matter haloes were identified using a halo finder based on the spherical overdensity method \citep{watson2013halo}. We only use dark matter haloes with a mass of $M_\mathrm{min}=10^9 M_\odot$ or higher. See \cite{dixon2016large} and \cite{giri2018bubble} for more details on the N-body simulation.

We model the ionising photon sources residing within the dark matter haloes as follows. The number of ionising photons ($\Dot{N}_{\gamma}$) emitted per unit time into the IGM is given as,
\begin{equation}
\label{eq:emissivity}
    \Dot{N}_{\gamma} = \zeta \frac{M}{m_{\mathrm{p}}}\frac{ \Omega_{\mathrm{b}}}{ \Omega_{\mathrm{m}}} \frac{1}{f_{\mathrm{coll}}} \frac{\mathrm{d}f_{\mathrm{coll}}}{\mathrm{d}t} \ ,
\end{equation}
where, $M$ and $m_{\rm p}$ are the dark matter halo and proton masses, respectively, and  $f_{\mathrm{coll}}$ is the collapsed fraction for $M\geq M_\mathrm{min}$.
We absorb all the complex processes, such ionising photon production and escape fraction, into the $\zeta$ parameter\footnote{Though this quantity can be mass dependent to match the high redshift galaxy observations, we assume it to be mass independent in this study \citep[see e.g.,][for discussions]{park2019inferring,giri202421}.}. For computational efficiency we combine dark matter halos which fall inside the same grid cell of our $250^3$ grid into one source of photons, equivalent to the sum of their masses. Therefore, each grid cell contains only one source.

\subsection{Spherical photon emission model} \label{sec:method_spherical}

We simulated the (re-)ionisation of the IGM using the fully numerical radiative transfer code \ccray{}\footnote{\url{https://github.com/garrelt/C2-Ray3Dm}} \citep{mellema2006c2,Hirling202pyc2ray}.
In the standard case, the code assumes that photons are emitted isotropically.
Our fiducial model has isotropic emission with emissivity parameter $\zeta=50$.
We model the impact of unresolved sinks in the IGM by setting a maximum distance that the ionising radiation can propagate, which we set to 70 cMpc during the EoR.  
See \cite{georgiev2024forest} for a comparison of several approaches to model unresolved sinks. This simulation is referred to with the label \sphere{} or simply as the fiducial model.

\begin{table}
    \centering
    \caption{List of model parameters considered in this work.}
     \begin{tabular}{||c c c c||} 
     \hline
     Name & Solid Angle ($\Theta_c$) 
     & Emission Direction & Boost Factor \\ 
     \hline\hline
     {\sc Sphere} & $4 \pi$ 
     & Spherical & 1.000 \\ 
    {\sc Cone-1} & 0.09$\pi$ 
    & Static Conical & 44.779 \\ 
    {\sc Cone-2} & 0.35$\pi$ 
    & Static Conical & 11.451 \\ 
    {\sc Cone-3} & 0.09$\pi$ 
    & Time-Evolving Conical & 44.779 \\ 
     \hline
     \end{tabular}
    \label{tab:models}
\end{table}

\subsection{Conical photon emission}\label{sec:method_conical}

The aim of this paper is to explore the impact of anisotropic emission from the sources of cosmic reionisation. Fig.~\ref{fig:One_Cone} illustrates a toy model containing two sources in a uniform background, both emitting an equal number of ionising photons into the IGM, following a black body spectrum with $T = 50,000$ K. The source in the bottom left corner emits ionising photons isotropically, creating a spherical ionised region. In three dimensions, this region forms a sphere. In contrast, the source in the top left corner emits photons directionally, primarily along the positive x-axis. 
This results in a conical ionised region with the source at the vertex and a hemispherical base with an opening angle of $\theta_c=0.35$\footnote{This shape is also referred to as a \textit{spherical sector} or \textit{spherical cone}.}.
We will refer to the conical-shaped region as the `cone' or `conical region' throughout this work.

The geometry of the escape of ionising photons from real galaxies is obviously more complicated than a sample cone shape as we assume here.
The direction of the emission from every source can be more complex, for example the photons might escape in multiple directions. High-resolution simulations of early galaxies \citep[e.g.,][]{razoumov2006escape,gnedin2008escape,trebitsch2017fluctuating} have shown that they could be linked to the galactic physics as well.
However, to establish if any measurable effect might be expected, a simple cone model such as the one described above should be sufficient. Our approach could be seen as a numerical experiment rather than a realistic modelling of how ionising photons escape from galaxies. In the following subsections, we describe the ionising sources used in our numerical simulations.

\begin{figure}
    \centering
    \includegraphics[width=0.5\columnwidth]{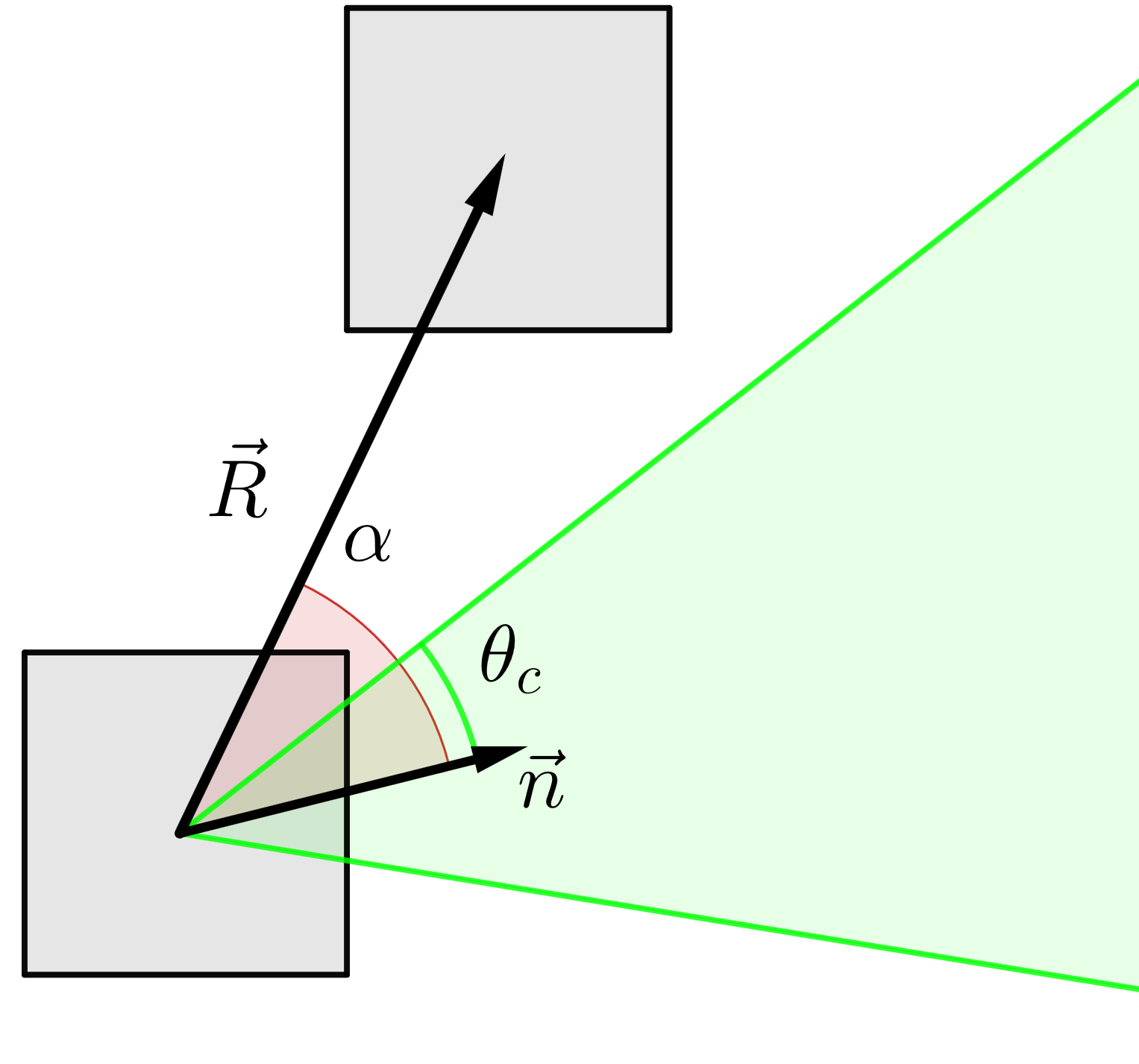}
    \caption{A two-dimensional illustration of the modified algorithm implemented in \ccray{} to model direction dependence of photon emission. The ionising source is located in the bottom left cell and the green-shaded region is a cross-section through the axis of the cone with unit vector $\vec{n}$ going through this, while $\alpha$ is the angle between the axis through the cone and the vector to any cell where impact of ionising photons is studied.
    }
    \label{fig:Diagram}
\end{figure}

We parameterise the opening of these cones through their solid angle, $\Theta_c = 2\pi(1-\cos\theta_c)$, where $\theta_c$ is the angle between the axis of the cone and its surface (see Fig.~\ref{fig:Diagram} for an illustration of the geometry used). We used the definition of solid angle and the standard expression for the area of a spherical cap, $\Theta_c=A/R^2=2\pi Rh/R^2=2\pi(1-\cos\theta_c)$. In this parameterisation, the spherical emission case corresponds to an opening angle of $\Theta_c=4\pi$.

We model the transfer of photons from the ionising sources using a (short-characteristic) ray tracing method \citep[see][for the detailed algorithm]{mellema2006c2,Hirling202pyc2ray}. In this algorithm, the total amount of photons are normally isotropically distributed in 3D. We modified this algorithm by adding a condition to check if the cells inside which rays are to be traced are within a certain emission cone. 
In Fig. \ref{fig:Diagram}, we show a 2D illustration of this method. If we assume a source to reside inside the lower left cell and the emission cone described with the green-shaded region that has an opening angle of $\theta_c$ and $\vec{n}$ is a unit vector through the axis of this cone. To check if any sink cell (top right) is inside the cone, we estimate the angle subtended at the cone axis, which can be given as,
\begin{equation}
    \alpha = \cos^{-1} \left(\frac{\vec{R}\cdot\vec{n}}{|\vec{R}||\vec{n}|}\right) \ ,
    \label{eq:alpha}
\end{equation}
where $\vec{R}$ is the vector from the source to the sink cell. If this angle is more than $\theta_c$, then the cell is considered beyond the simulation's hard barrier and the photoionisation rate for this cell is set to zero. 
This will stop the rays from being traced in this sink cell.

Due to this approach, fewer photons reach the IGM than in the isotropic case as photons emitted outside the cones are assumed to be completely absorbed in the interstellar medium and not allowed to leave the galaxy. In order to preserve the total number of ionising photons on the grid and therefore produce similar reionisation histories, we boost the sources by a factor $B$. 
This boost parameter is the ratio of the total solid angle and the angle subtended by the cone, which is given as
\begin{equation}
    B = \frac{4\pi}{2\pi(1-\cos{\theta_c})} = \frac{4\pi}{\Theta_c} \ .
	\label{eq:boost}
\end{equation}
Another way to view this boost factor is to say that in the spherical emission case, the assumed escape fraction is $1/B$ and in the conical emission case, the escape fraction is 1 inside the cone and 0 outside it. The effect is that both types of sources release equal number of photons into the IGM, which allows us to isolate the effects of anisotropy.

We list the model parameter values for our simulations in Table~\ref{tab:models}. 
We study the impact of the opening angle on the reionisation geometry by considering models with $\theta_c=0.3$ and 0.6, or $\Theta_c = $ 0.09$\pi$ ({\sc Cone-1}) and 0.35$\pi$ ({\sc Cone-2}), respectively \citep[see][for a discussion on opening angles]{Roy2015}.
In these two cases, the directions of the cones are randomly set at the time of appearance of the dark matter halo and kept constant after that. Below we will also refer to these two cases as 'static'. The values for the boost factor $B$ show that the equivalent values of the escape fraction ($1/B$) for our cases are 9 and 2 per cent, values which are in the range claimed for both observed and modelled galaxies \citep[e.g.,][]{gnedin2008escape,mitra2015cosmic,grazian2016lyman,rutkowski2016lyman}. We also considered a scenario (\conethree{}) where the cone opening angle is identical to the one in the {\sc Cone-1} model but the direction of each cone was randomly changed every 11.5 Myrs (the time between outputs from our N-body simulation).
The goal of this model was to test whether changes in the direction of the cone could smear out the effect of the conical emissions. 

As pointed out above, in our simulation each grid cell only contains one source, even though it could contain multiple halos. In Appendix~\ref{sec:clustered_sources_emission}, we discuss this unresolved clustering of sources within a grid cell of size $\sim 1 h^{-1}$~Mpc and how this affects the anisotropy of photon emission. 

\section{Results}\label{sec:results}
\begin{figure*}
    \centering
    \includegraphics[width=1.0\linewidth]{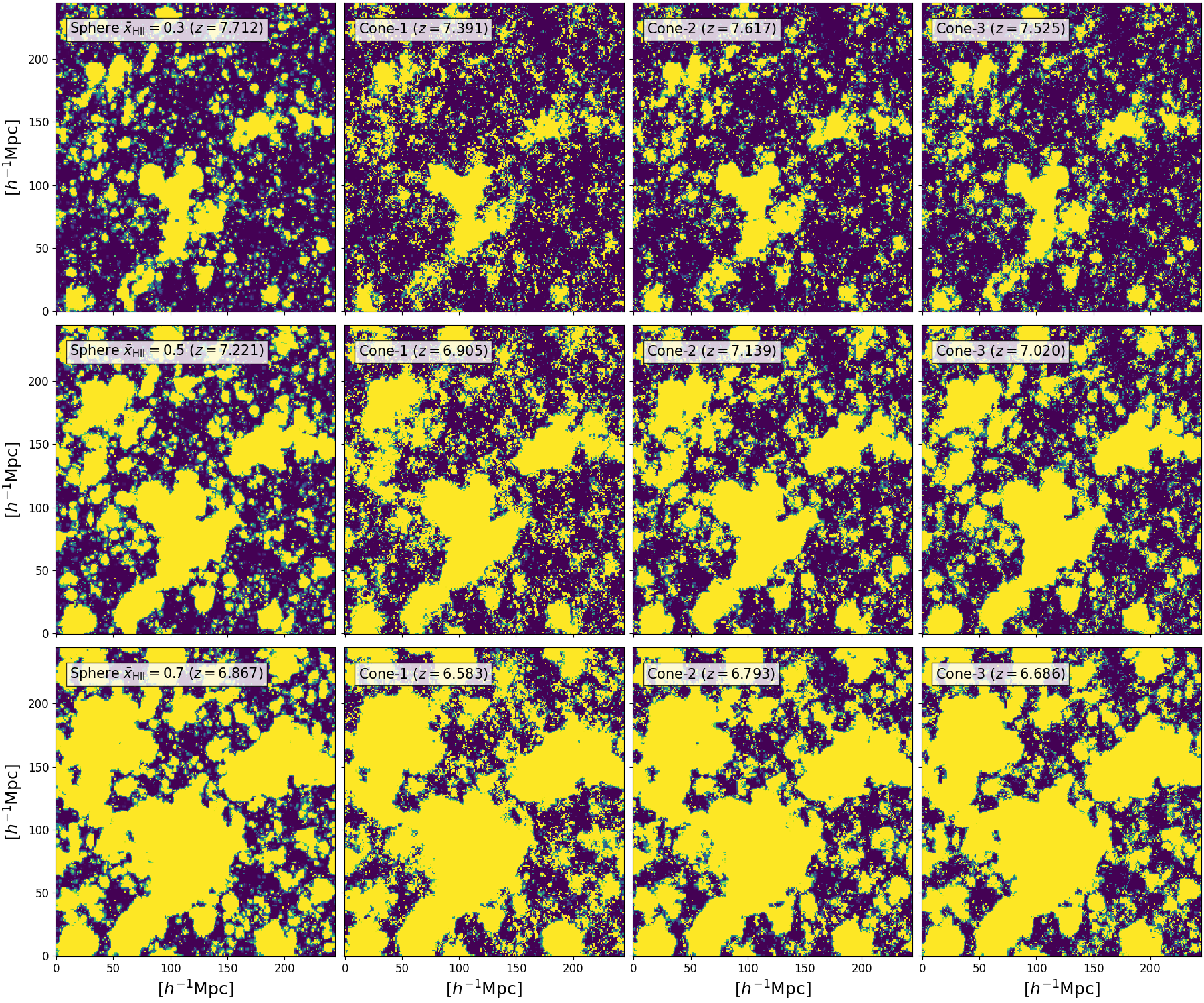}
    \caption{Slices of the simulations at the same ionisation fraction in each row. Bright yellow and dark blue pixels correspond to ionised and neutral intergalactic medium, respectively.
    Far left ({\sc Sphere}): The fiducial isotropic case that is being used for comparisons. 
    Centre left ({\sc Cone-1}): The conical case with static directions of cones for every source with opening angle $\Theta_c=0.09\pi$.
    Centre right ({\sc Cone-2}): The conical case with static directions of cones for every source with opening angle $\Theta_c=0.35\pi$.
    Far right ({\sc Cone-3}): The conical case with cones of opening angle $\Theta_c=0.09\pi$, where the direction of the cone of every source is randomly changed every time step.
    In all cases, the number of separate ionised regions increases when we restrict the emission of ionising photons to cones.
    } 
    \label{fig:Slices}
\end{figure*}

We now present the results studying the impact of direction dependence of ionising photon leakage from the suite of reionisation simulations produced in this work. For a first visual impression we show in Fig.~\ref{fig:Slices} ionisation fraction slices from all four simulations at three different stages of reionisation. The left column shows the {\sc  Sphere} model and the second, third and rightmost columns {\sc Cone-1}, {\sc Cone-2} and {\sc Cone-3}, respectively. These slices show that each of these models produces slightly different morphologies for the ionised
regions. The majority of differences are observed at small scales. As the opening angle increases or the direction is regularly randomised the results become closer to the isotropic case, as could be expected. 
In the rest of this section we analyse our results in more detail. 
\subsection{Reionisation History}\label{sec:results_history}
\begin{figure}
    \centering
    \includegraphics[width=1.0\linewidth]{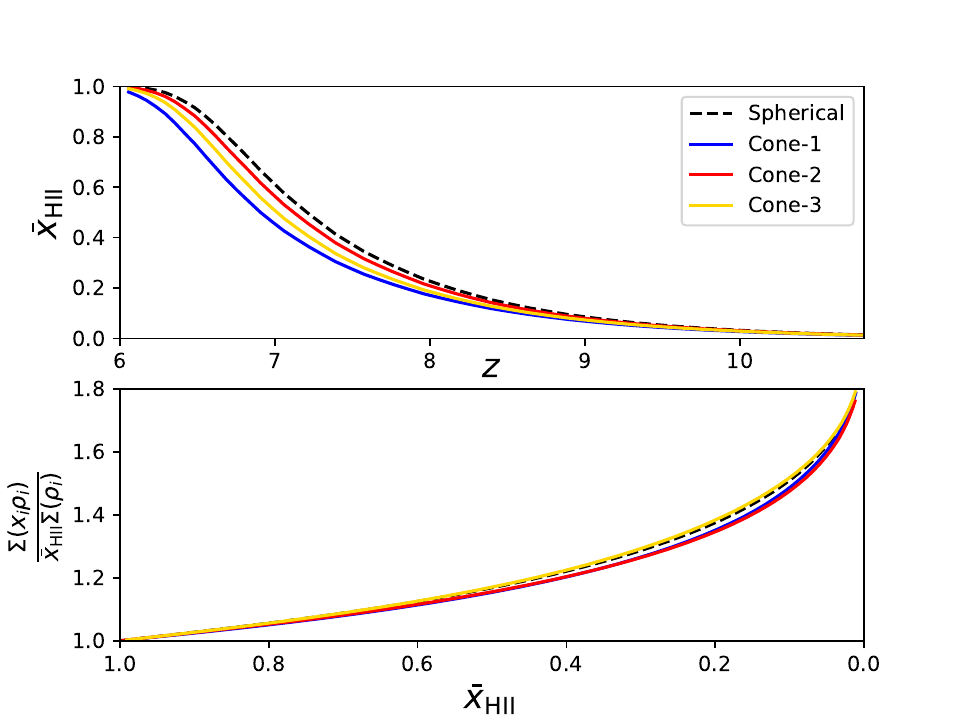}
    \caption{Top Panel: The evolution of the volume-averaged mean ionisation fraction $\bar{x}_\mathrm{HII}$ of the simulations developed in this work. We observe that reionisation is delayed when ionising photons leak through small channels.
    Bottom Panel: The ratio between the mass-weighted ionisation fraction and the volume weighted fraction.
    In all the models, the nature of reionisation remains \textit{inside-out}.
    }
    \label{fig:Reionization}
\end{figure}

We first consider the reionisation history of our simulations. 
In Fig.~\ref{fig:Reionization}, we show the volume-averaged ionisation fraction (top panel). 
Although we adjusted the number of ionising photons emitted by the sources in the conical models through a boost factor (Eq.~\ref{eq:boost}) to match the \sphere{} model, reionisation proceeds more slowly in the anisotropic models. The \coneone{} model, which has the smallest conical emission channels, has the slowest history among all the models. 
The delay in the reionisation history is due to resolution effects, as is explained in Appendix \ref{sec:resolution_b}.

The \conetwo{} model uses a larger conical opening angle, compared to that of \coneone{}, and consequently the reionisation history is closer to the \sphere{} model. As expected, the reionisation history of \conethree{} lies between \coneone{} and \conetwo{} as this model has the same opening angle as \coneone{}, but with the direction changing over time. This change in direction reduces the anisotropy in emission when averaged over time. In the later part of the paper, we will compare the models at the same volumed-averaged ionisation fraction.

The bottom panel of Fig.~\ref{fig:Reionization} presents the ratio of the mass- and volume-averaged ionisation fractions. The ratio is always greater than unity in all our models (bottom panel), indicating that the reionisation process remains on average \textit{inside-out} throughout the evolution. 
Note, however, that for \coneone{} and \conetwo{} models these values are noticeably smaller, particularly during the middle stages of reionisation.
The reason for this is that sources tend to form in high-density regions in the knots of the cosmic web of structures.
In the spherical case, all surrounding space must be ionised before radiation can reach cells further away, and these tend to be low-density cells.
In the static anisotropic cases, the radiation is concentrated in a cone and if that cone does not align with any filaments, there are high-density cells that are missed, while some of the low-density voids, which take up most of the volume, are ionised earlier.  

Furthermore, \conethree{} model has a noticeably greater ratio than the other two codes, close to the \sphere{} model. 
This is due to the randomly changing direction of the cones, resulting in nearby, high-density cells, previously neutral, being the focus of all of a source's radiation.
While the reionisation history is altered by the anisotropic leakage of ionising photons, the effect on the mean reionisation history is degenerate with other astrophysical parameters, such as the ionising photon production efficiency and minimum halo mass required for hosting sources. 

In the next sections, we focus on the properties of the ionised regions, as well as on several other observables which could be expected to be more sensitive to the directionality of the photon escape. 

\subsection{Ionised Bubble Sizes}\label{sec:results_bsd}
The snapshots in Fig.~\ref{fig:Slices} clearly show that the sizes and percolation of the ionised regions for the anisotropic cases differ from the \sphere{} model. Therefore our next aim is to quantify this by considering the \HII region size distributions determined using the Mean-Free-Path \citep{Mesinger2007} and Friends-of-Friends \citep{Iliev2006} methods. See \citet{giri2018bubble} for a comparison of these methods. 
\begin{figure*}
    \centering
\includegraphics[width=0.95\linewidth]{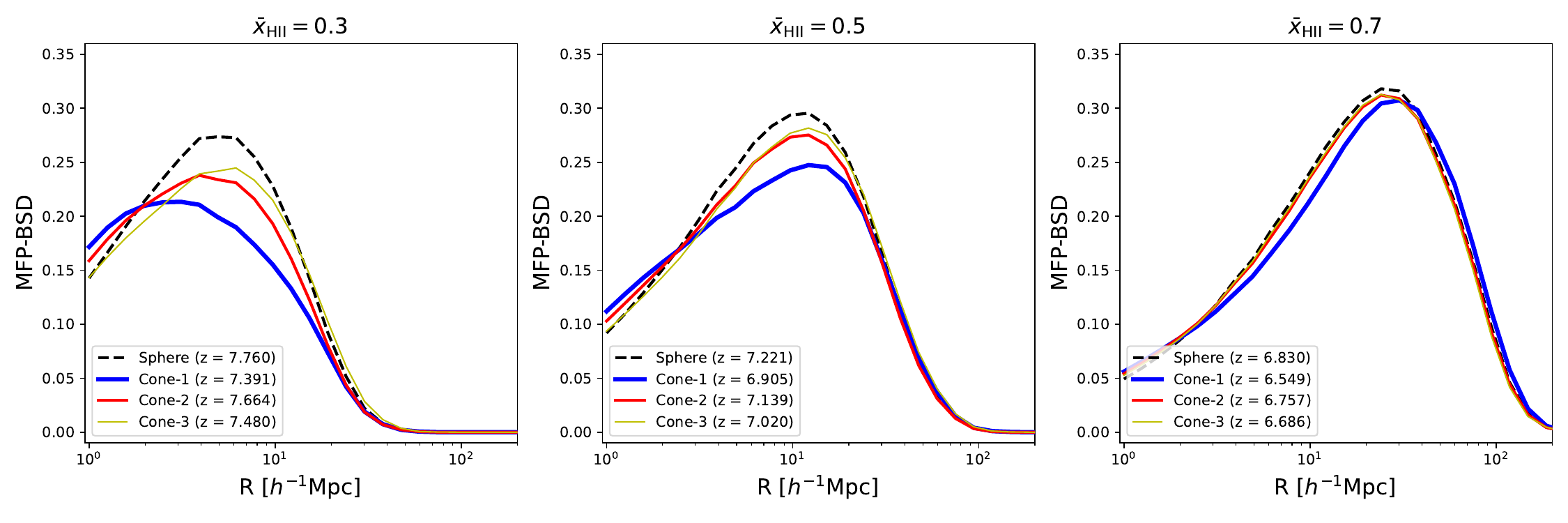}\\  
\includegraphics[width=0.95\linewidth]{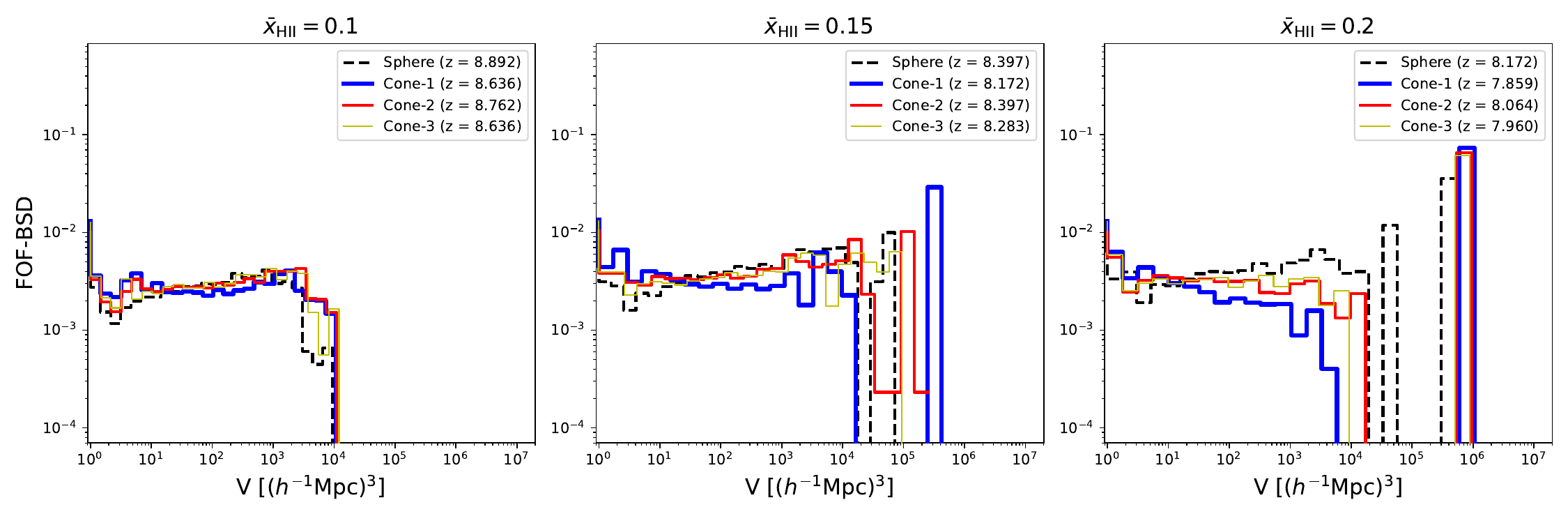}\\  
\includegraphics[width=0.95\linewidth]{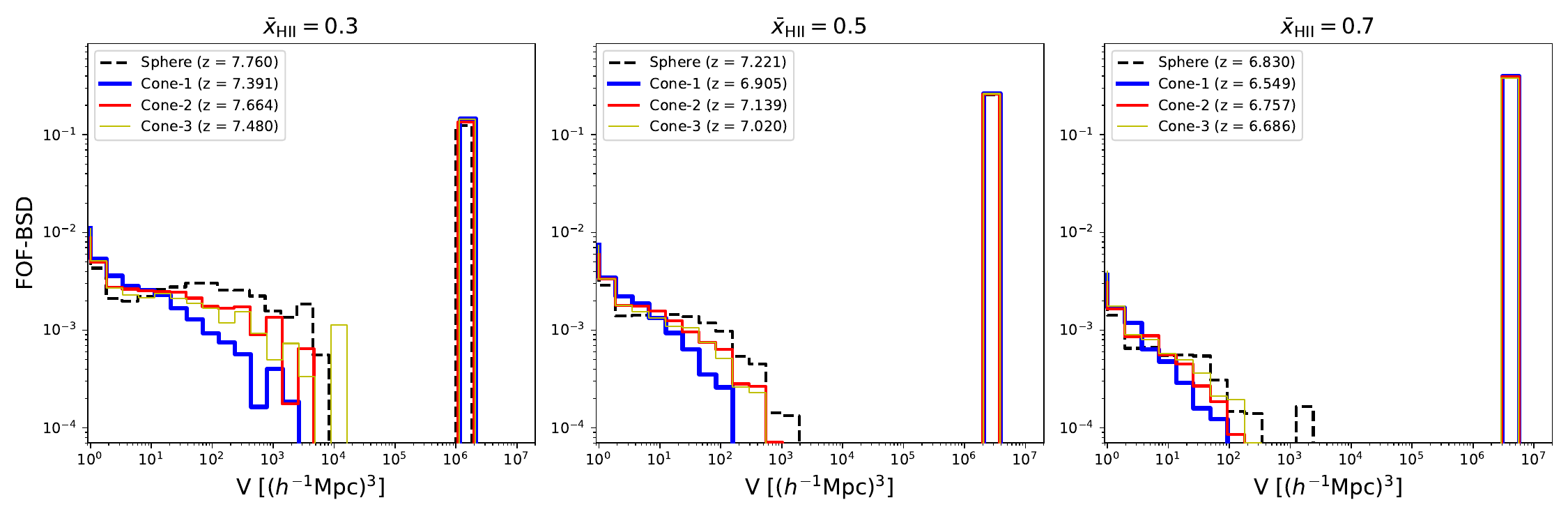}\\  

    \caption{The ionised bubble size distribution (BSD) estimated using the MFP (top panels) and FOF (middle and bottom panels) methods. 
    The top panels show the MFP BSDs for, from left to right, $\bar{x}_\mathrm{HII} = $ 0.3, 0.5, and 0.7. These results show that the typical sizes of ionised regions in the conical models are smaller than those in the \sphere{} model during the early stages of reionisation but are biased toward larger sizes during later times.
    The middle panels show the FOF BSDs for $\bar{x}_\mathrm{HII} = $ 0.1, 0.15, and 0.2 and the bottom panels for $\bar{x}_\mathrm{HII} = $ 0.3, 0.5, and 0.7. 
    The FOF BSDs present the connectivity of the ionised regions and show a significant difference between the models during the early stages ($\bar{x}_\mathrm{HII}\lesssim 0.5$).}
    \label{fig:BSDs}
\end{figure*}

\subsubsection{Mean-free-path bubble size distributions}\label{sec:results_mfp_bsd}

In this subsection, we investigate the effect anisotropic emission has on the size distribution of ionised regions during reionisation using the Mean-Free-Path Bubble-Size-Distribution (MFP-BSD) estimation method described in \citet{Mesinger2007}. 
In this method, we keep track of the length of a ray from an ionised cell ($x_\mathrm{HII}\ge 0.5$) to the first encountered neutral cell ($x_\mathrm{HII}< 0.5$).
The process is iterated for a large number ($\sim 10^7$) of rays, randomly initiated across different ionised regions, culminating in an overall, and normalised, distribution of ray lengths. 
The resulting distributions for our suite of simulations are shown in the top panels of Fig.~\ref{fig:BSDs}, compared at a fixed volume-averaged ionisation fraction. 

During the early stages of reionisation ($\overline{x}_{\mathrm{HII}} \approx 0.3$; top-left panel), the anisotropic emission has the most pronounced effect on the bubble size distribution. 
The fiducial \sphere{} model reaches this point at a slightly higher redshift than the anisotropic models, 
and has the narrowest size distribution, with a peak value of $R \approx 6$ $h^{-1}$Mpc. 
The \conetwo{} and \conethree{} models yield very similar distributions,
which are wider and peak at somewhat lower sizes, $R\approx 4-5\,$ $h^{-1}$Mpc. As reionization is slightly delayed in these models, they are reached slightly later, by $\Delta z \approx 0.1-0.2$ compared to our fiducial model. This delay in reionisation and shift towards smaller ionised patches and a wider size distribution becomes more pronounced in the \coneone{} model, where the smallest opening angle further delays reionisation by $\Delta z \approx 0.3$, reaching a peak at only $R\sim3\,$$h^{-1}$Mpc. 
Overall, we see that the differences in shape between the conical and spherical regions directly impact the MFP-BSD distribution width. 
The distance from the centre to the surface of a spherical region is more isotropic than that of a cone of the same volume, which yields a more uniform distribution peaked at higher sizes (dashed model). 

Interestingly, for the \conethree{} model, introducing the variable cone directions produces an MFP-BSD distribution closer to that of the \conetwo{} model, despite having the same opening angle as \coneone{}.
The reason for this is likely because due to the directions of the cones changing, the resulting bubbles become more spherical and begin overlapping earlier, something shown below in Fig.\ref{fig:BSDs}, the FOF-BSD. As the sources are strongly clustered, the \conethree{} bubbles will be closer to the isotropic case. 

As the EoR progresses, the aforementioned features become less pronounced, with all models having a similar MFP-BSD peak of $R = 10~h^{-1}$ Mpc at the midpoint (top-middle panel of Fig.~\ref{fig:BSDs}). This distribution for anisotropic cases remain wider, if less so than early on. The distributions become nearly identical by $\overline{x}_{\mathrm{HII}} \approx 0.7$, with only the most anisotropic case (\coneone{}) showing a slightly different distribution, with fewer \HII regions at sizes smaller than the peak.

\subsubsection{Friends-of-Friends bubble size distributions}\label{sec:results_fof_bsd}

Another method we used to quantify bubble sizes is the Friends-of-Friends Bubble-Size-Distribution (FOF-BSD) method \citep[][]{Iliev2006,giri2018bubble}. 
In this method, if two cells that are both ionised ($x_\mathrm{HII}> 0.5$) and share a face, they are considered ``friends'' (see Appendix~\ref{sec:ionisation_limit} for a discussion on the impact of the choice of the ionisation threshold). 
Beginning the process for a given cell, the method finds all the friend cells and then counts all of their friend cells (hence ``friends of friends'') ad infinitum. In practice, this allows us to identify large connected irregularly shaped ionised regions. One crucial advantage of the FOF-BSD compared to the MFP-BSD comes from the fact that each bin of the FOF-PDF maps the occurrence of continuous, topologically-connected regions of ionised cells of volume $V$, physically separated from the other distinct regions by intervening neutral hydrogen.  This method is also sensitive to the percolation process, whereby ionised patches around individual sources or clustered groups of sources merge together into a singular connected structure  \citep[see][]{Furlanetto2016}. 

We examine the FOF-BSD at a range of $\overline{x}_{\mathrm{HII}} \approx 0.1 - 0.2$, presented at the middle row of Fig.~\ref{fig:BSDs}.
Initially, is FOF-BSD is relativity similar for all simulations (top left column), with the exception of the \sphere{} case where the fewer ionised structures between bins of $10^3$ $(h^{-1}\mathrm{Mpc})^{3}$ > V < 10$^4$ $(h^{-1}\mathrm{Mpc})^{3}$ compared to the anisotropic models. Examined at a fixed global neutral fraction, the ionised sources in the \sphere{} model are fewer and more biased to the large scale density field, initially resulting in fewer overlapping structures. Conversely, the anisotropic models reach the same global neutral fraction at a lower redshift of $\Delta z \approx 0.1$. The larger number of ionising sources results in more isolated regions with small sizes, while the narrow opening angles in the anisotropic emission model promote the formation of large ionised structures. These effects become most pronounced in the middle right panel when all models form the largest connected ionised region of the order of $V \approx 10^{6}$ $(h^{-1}\mathrm{Mpc})^{3}$. This large ionised cluster \citep[as discussed in e.g.,][]{Iliev2006,giri2018bubble} is fundamentally linked to the process of percolation \citep{Furlanetto2016}. In essence, the formation and overlap of the first bubbles occur along the filaments of the cosmic web, creating a long ionised region with a complex twisting and winding geometry, which can span the size of the simulation box \citep[][]{Iliev2006,giri2021betti}. Note that, in all FoF-BSD panels of Fig.~\ref{fig:BSDs}, the volume of this region increases with reionisation as more and more isolated ionised regions merge with it. 

We can track the formation and evolution of the percolation cluster by examining the FoF-BSD panels where $\overline{x}_{\mathrm{HII}} = 0.15,0.2, 0.3$ in Fig.~\ref{fig:BSDs} (middle and last panels of the second row, and the first panel of the third row). Here, the role of the anisotropy, can be clearly seen to influence the initial size of the largest FoF cluster, which in the case of the \sphere{} model is slower to percolate compared to the anisotropic models (middle right panel). In this period the \sphere{} case has a larger abundance of regions with V > $10^3$ $(h^{-1}\mathrm{Mpc})^{3}$, hinting that while large ionised regions are present in the simulation, the spherical emission does not result in a higher chance of overlap. In fact, it would appear that the narrow the opening angle the cone is, the quicker the formation and the growth of the largest ionised FoF regions. 

Throughout the EoR individual clusters with volumes of $V < 10$ $(h^{-1}\mathrm{Mpc})^{3}$, are more numerous for the anisotropic emission models. 
The narrower the cone, the more pronounced this effect is, consistent with what is observed from short path lengths in the MFP-BSD (top panels of Fig.~\ref{fig:BSDs}). 
This can be intuitively confirmed by observing \coneone{} and \sphere{} in the upper panels of Fig.~\ref{fig:Slices}, noting the grainy structure produced by the isolated conical shapes. 
On the other hand, clusters with $V > 10$ $(h^{-1}\mathrm{Mpc})^{3}$, except for the percolation cluster, are more numerous the wider or more spherical the emission. 
The \sphere{} model has a flatter distribution compared to the conical models, which have a PDF decreasing with volume, which is steeper for narrower cones. 
This is consistent with the MFP-BSD results for the \coneone{} case, where the number of ionised lines-of-sight below the peak of the distribution is always smaller compared to the other models. 
The flat shape of the FOF-BSD observed in the \sphere{} case can be attributed to $x_{\mathrm{HII}}$ bubbles/clusters located in voids, whose isotropic growth results in a more compact cluster, less likely to overlap with a neighbouring cluster.

\subsection{Observational signatures at the redshifted 21-cm line}

\begin{figure*}
    \centering
    \includegraphics[width=1.0\linewidth]{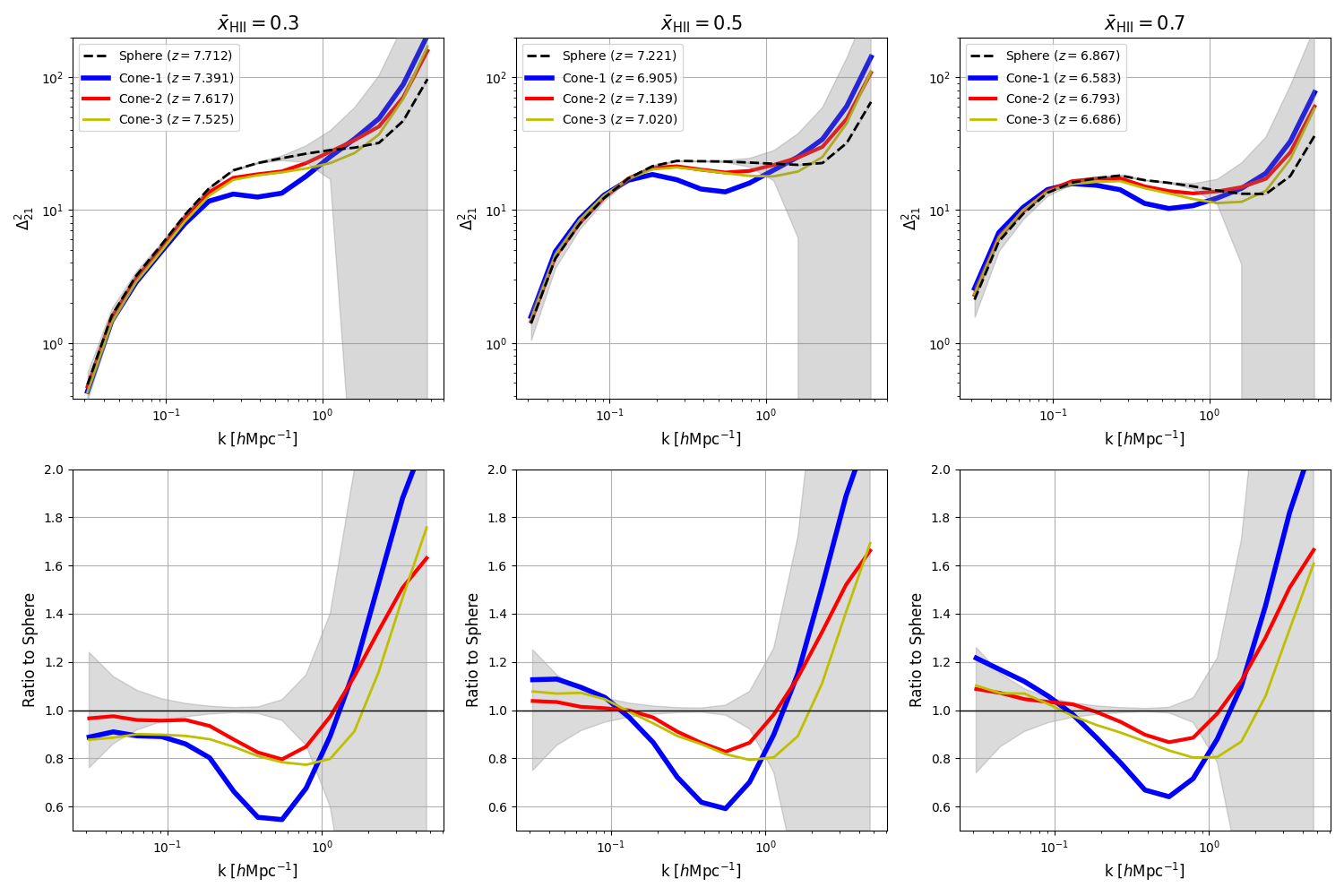}
    \caption{The spherically-averaged power spectrum of the 21-cm signal $\Delta^2_\mathrm{21}$ (top panels) and their ratio with respect to the \sphere{} model (bottom panels) at mean ionisation fraction $\bar{x}_\mathrm{HII}=0.3$ (left), 0.5 (middle), and 0.7 (right). The power is suppressed at scales $k\sim 0.1-1h\,\mathrm{Mpc}^{-1}$ in the conical emission models compared to the spherical case. The shaded regions shows the error due to thermal noise expected from 1000 hour observation with the SKA-Low \citep[see][]{schneider2023cosmological}.
    }
    \label{fig:PS_clean}
\end{figure*}

The redshifted 21-cm signal probes the distribution of the neutral hydrogen. Radio interferometers observe the 21-cm differential brightness temperature, given by \citep[e.g.][]{mellema2013reionization}
\begin{align} \label{eq::dTb} &\delta T_{\mathrm{b}} (\mathbf{r}, z) = T_{0} (z) x_{\rm HI}(\mathbf{r}) (1+\delta_{b}(\mathbf{r})) \bigg(\frac{T_{\mathrm{s}}(\mathbf{r},z) -T_{\mathrm{CMB}}(z) }{T_{\mathrm{s}}(\mathbf{r},z) }\bigg) , \\ 
&T_0 (z) \approx 27 \left(\frac{1+z}{10}\right)^{\frac{1}{2}} \left( \frac{\Omega_{\mathrm{b}}}{0.044} \frac{h}{0.7} \right) \left( \frac{\Omega_{\mathrm{m}}}{0.27} \right)^{-\frac{1}{2}} \bigg( \frac{1-Y_{\mathrm{p}}}{1-0.248} \bigg) ~\mathrm{mK},
\end{align}
where $\delta_b$ is the mass density contrast, $x_{\rm HI}$ is the neutral fraction, $Y_\mathrm{p}$ is the primordial helium abundance by mass, while $T_{s}$ and $T_{\mathrm{CMB}}$ are the spin and CMB temperature respectively. We explore the case of $T_{\mathrm{s}}\gg T_{\mathrm{CMB}}$ where the gas in the IGM is heated by the first ionising sources and we do not consider the effects of redshift space distortions, which also makes the signal anisotropic \citep[e.g.][]{jensen2013probing,ross2021redshift}. 
However, we aim to study whether anisotropic photon emission independently introduces any anisotropy in Sec.~\ref{sec:results_rmu}.

\subsubsection{21-cm Power Spectrum}\label{sec:results_ps}
\label{sec:21cm_ps}

In this section, we present the dimensionless power spectrum 
at a certain wavenumber $k$ as $\Delta^{2}_{\rm 21}(k)=k^{3} P(k)/(2\pi^{2})$, where $P(k)$ is the spherically-averaged power spectrum. Figure \ref{fig:PS_clean} shows the 21-cm power spectrum for all the models we consider in the top panel, and on the bottom panel we plot the ratio of each anisotropic emission model against the fiducial \sphere{} case. The grey bands highlight the uncertainty on the measurement as we take into account thermal noise (rightward grey band) and cosmic sample variance (leftward gray band), i.e. for an observation with SKA-Low for approximately 1000 hours. We followed the method described in \citet{giri2022imprints} and \citet{schneider2023cosmological} to model this telescope effect. For a detailed discussion about the assumed telescope properties and observation strategy, we refer to section~\textit{V} in \citet{giri2022imprints}.
Note that in the regime where $k<0.1\,h\,\mathrm{Mpc}^{-1}$ the 21-cm power spectrum is most prone to uncertainty due to residual foreground contamination. 

The effect of the anisotropic emission on the 21-cm power spectrum is widely pronounced across all $k$-scales during the EoR. Compared to the fiducial \sphere{} model at a fixed ionisation fraction, the small-scale 21-cm power spectrum ($k\ge 0.8\,h\,\mathrm{Mpc}^{-1}$) has a higher amplitude the narrower the cone is. Such $k$-scales correspond to ionised regions with approximate sizes of $R \le 2 \pi /k = 7.8 \, h^{-1}$ Mpc. Visually comparing the ionisation slices at the top panel of Fig.~\ref{fig:Slices}, we see that the excess power is indeed associated with the small-scale structure induced by the anisotropic emission. This is consistent with what we observe for the distributions of the MFP-BSD and FOF-BSD methods in Fig.~\ref{fig:BSDs}. Naturally, the effect becomes slightly less pronounced with the progression of the EoR (and, therefore, at lower redshifts) as ionised structures overlap with the progress of the EoR. However, while the 21-cm power spectrum at high-$k$ values differs in amplitude and evolution between the conical and spherical models, finer differences between the anisotropic emission models are difficult to disentangle. Moreover, since the thermal noise of the 21-cm power spectrum increases with $\Delta^{2}_{\mathrm{noise}} \propto k^{3/2}$ as well as with redshift \citep[see eq. 17 and 18 of][]{Georgiev2024}, exploring this regime requires larger integration time to discern between different anisotropic emission models. 

The large-scale 21-cm power spectrum ($k\leq 0.1\,h\,\mathrm{Mpc}^{-1}$) is quite similar between all types of emission, however, differences in the overall amplitude of the low-$k$ power spectrum of around 5-10 per cent can be seen for the \coneone{}, \conetwo{}, and \conethree{} models after the midpoint of reionisation. This implies that for the anisotropic emission models, ultraviolet photons emitted in tighter bunches due to the narrowing cone can propagate further across the ionised region, resulting in more rapid growth of the largest ionised structures. Indications of this can be noted when comparing the amplitude of \coneone{} against the \sphere{} model for the large ionisation cluster in the FoF distributions (around $V \approx 10^{6} \, (h^{-1}$Mpc$)^{3}$). However, as seen in the bottom panel of Fig. \ref{fig:PS_clean} for low-$k$ values, discerning the mode of emission of the ionising sources becomes a difficult task when considering the uncertainty on large-scale 21-cm power spectrum due to the limited field of view of radio interferometers.

The most interesting results are seen in the ``intermediate'' region of the 21-cm power spectrum (for $0.1 < k < 1\,h\,\mathrm{Mpc}^{-1}$), which is the primary target for most radio interferometers due to the lower uncertainty caused by sample variance and thermal noise. In the bottom-left panel of Fig. \ref{fig:PS_clean}, we see a decrease in the amplitude of the 21-cm power spectrum at $k \approx 0.4\,h\,\mathrm{Mpc}^{-1}$ by a decrease of 40 per cent (a factor of 2 of the 21-cm power spectrum) for the \coneone{} model compared to the \sphere{} model. While this difference diminishes over time, a lower amplitude of 30 per cent (a factor of 1.5 of the 21-cm power spectrum) is still observed at $\bar{x}_{\mathrm{HII}} = 0.7$. Meanwhile, in the case of a wide cone, \conetwo{}, or a varying narrower cone, the difference in amplitude is significantly reduced (around a factor of 10-15 per cent lower than in the spherical case) and also slowly diminishes with the EoR. Moreover, the difference in amplitude between the anisotropic models and the fiducial case has a $k$-scale dependency, which becomes more pronounced as $k$ increases until a turnover stage is reached, after which the anisotropic models exhibit more power. 
This effect is directly tied to the principal differences between conical and spherical structures of the ionised regions, which we discuss in Appendix~\ref{sec:simple_case} in greater detail. 

\subsubsection{21-cm Anisotropy Ratio}\label{sec:results_rmu}

\begin{figure*}
    \centering
    \includegraphics[width=1.0\linewidth]{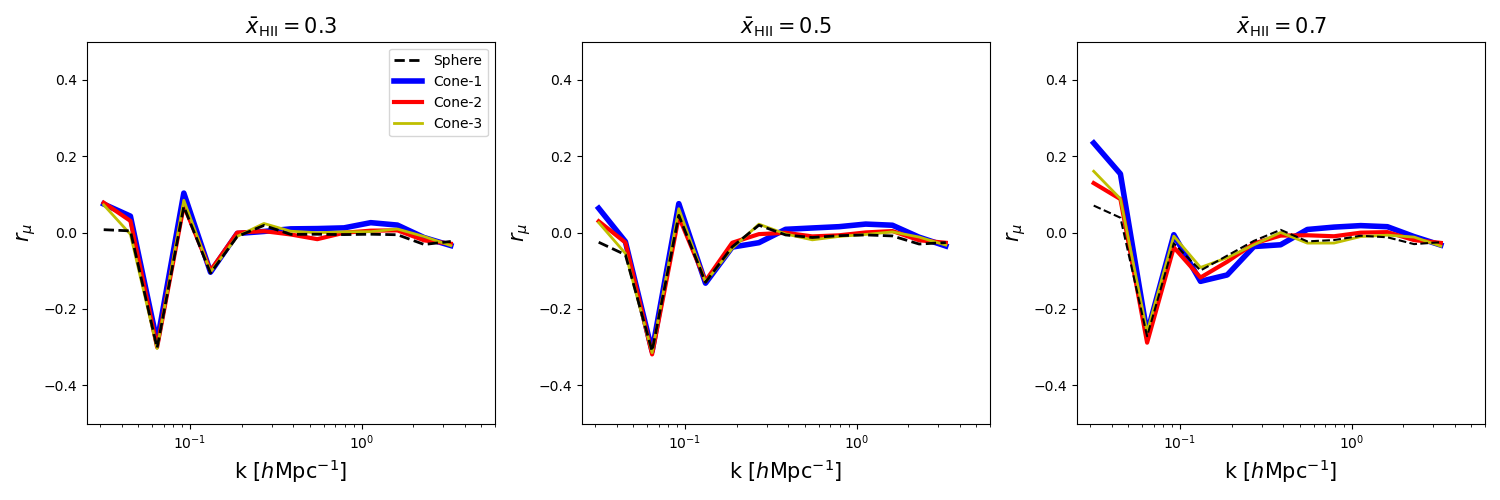}
    \caption{The anisotropy ratio compared at mean ionisation fraction $\bar{x}_\mathrm{HII}=0.3$ (left), 0.5 (middle) and 0.7 (right). The four simulation models are represented with different lines listed in the legend. The values are very close to zero indicating absence of any anisotropy in the 21-cm signal from our simulations.}
    \label{fig:rmu_clean}
\end{figure*}

In order to test if the direction dependence of the photon emission introduces any anisotropy in the 21-cm signal, we calculate the anistropy ratio $r_\mu$. This quantity is defined as \citep[e.g.,][]{fialkov2015reconstructing,ross2021redshift},
\begin{equation}
\label{eq:anisotropy_ratio}
    r_\mu(k) = \frac{\langle \Delta^2_\mathrm{21}(k,|\mu|>0.5)\rangle}{\langle \Delta^2_\mathrm{21}(k,|\mu|<0.5)\rangle} - 1 \ ,
\end{equation}
where $\mu$ is the cosine of the angle subtended to the line-of-sight direction. $r_\mu$ is a simpler test of anistropy compared to the full decomposition into the polynomials of $\mu$ \citep[see e.g.,][]{barkana2005method,jensen2013probing}. As we have not included redshift-space distortions in the 21-cm signal \citep[e.g.,][]{jensen2013probing,ross2021redshift} any deviation of $r_\mu$ from zero corresponds to anisotropy in the signal due to the conical emission.

In Fig.~\ref{fig:rmu_clean}, we present the $r_\mu(k)$ estimated for our models at three different reionisation epochs. We observe that the curves are very close to zero at all scales. This behaviour contrasts with the simple scenario of a single ionising source, as discussed in Appendix~\ref{sec:simple_case}. While individual sources emitting photons anisotropically could produce anisotropic ionised bubbles, the cumulative effect of many sources mitigates this anisotropy.
The fluctuations in the curves at large scales ($k<0.2~h\,\mathrm{Mpc}^{-1}$) are due to sample variance. Therefore, these plots indicate that conical emission does not introduce any significant anisotropy at any scale. As our simple metric (Eq.~\ref{eq:anisotropy_ratio}) does not show any sign of anisotropy, we decided not perform any advanced analysis exploring anisotropy, such as $\mu$-dependent polynomial expansion \citep[e.g.,][]{barkana2005method}.

\section{Conclusions} \label{sec:conclusion}

In this work, we tested the impact of direction-dependent leakage of ionising photons into the intergalactic medium (IGM) during the epoch of reionisation (EoR). 
We modified our state-of-the-art cosmological reionisation simulation code, \ccray{}, to enable it to handle anisotropic emission of ionising photons, modelled as cone-shaped channels with adjustable opening angle and directions. We studied the effects of this using a suite of numerical reionisation simulations in volumes of $244~h^{-1}$ Mpc per side. In the anisotropic scenarios, the sources created ionised regions resembling cones with hemispherical bases. Reionisation of the IGM was driven by a large number  of (mostly weak) ionising sources. 
The collective effect of these sources might be expected to average out the individual source emission anisotropies, however our simulations revealed that the directional leakage of ionising photons left a number of imprints in the reionization morphology, some of which might be observable in the redshifted 21-cm signal.  Specifically, we found the following:
\begin{enumerate}
    \item We studied the size distribution of ionised bubbles in our simulations based on both the MFP-BSD and FOF methods. Both methods find more small, isolated \HII regions in the anisotropic cases compared to the fiducial distribution. This trend is stronger for smaller opening angles and during the early stages of reionisation (Fig.~\ref{fig:BSDs})
    and is weakened for randomised cone directions.
The role of the anisotropy of the emission is thus twofold. First, the principal shape of the cone is anisotropic, resulting in a MFP-BSD distribution shifted towards smaller scales compared to that of a sphere (see Appendix \ref{sec:simple_case} for an example). Second, the concentration of the ionising photons into smaller channels delays the merging of individual ionised regions early on. Beyond the middle stages of reionisation, these differences are reduced due to a significant increase in overlapping ionised structures.
\item In terms of potential direct observability, we show that the anisotropic emission from sources suppresses the 21-cm signal power spectrum by between 10 and 40\% at scales between $k\approx 0.1$ and $1$ $ h\,\mathrm{Mpc}^{-1}$, which are exactly targeted by the current radio experiments (see Fig.~\ref{fig:PS_clean}).
\item We find that the anisotropic emission from many ionising sources in a cosmological volume does not introduce any additional anisotropy in the 21-cm signal. This is because while the basic conic shape is indeed anisotropic, the large number of random directions as well as the substantial overlap of ionized regions erase this feature even in the early phases of reionisation (see Appendix \ref{sec:clustered_sources_emission} for a discussion).
\end{enumerate}

It is important to reiterate that our implementation of anisotropy is very simplified. Thus, our study serves only as a first indication that neglecting anisotropy may introduce certain differences compared to the isotropic case. For the 21-cm power spectrum we find a significant difference which could significantly impact the inferred parameters from interpreting the measured power spectrum. Therefore, our estimates provide a first rough upper estimate of the modelling uncertainty introduced by assuming isotropic emission, although more work is clearly needed to obtain better estimates.

In the future, we plan to develop and refine our model by connecting the direction and opening angles of the anisotropic photon emission to the properties of the dark matter haloes hosting the sources, such as their angular momentum vectors, as well as developing a method to more accurately account for the resolution effects in the cone model.

\section*{Acknowledgements}
TPS would like to thank the Department of Astronomy in Stockholm University for their assistance during the long-term attachment (LTA) period there. 
We acknowledge the useful discussions with Andrei Mesinger during the ``Understanding the Epoch of Reionization II'' workshop at the Sexten Center for Astrophysics.
GM’s research is supported by the Swedish Research Council project grant 2020-04691\_VR. SKG is supported by NWO grant number \textsc{OCENW.M.22.307}. Nordita is supported in part by NordForsk. 
The authors gratefully acknowledge the Gauss Centre for Supercomputing e.V.\footnote{\url{www.gauss-centre.eu}} for funding this project by providing computing time on the GCS Supercomputer JUWELS at J{\"u}lich Supercomputing Centre (JSC).
Other results were obtained on resources provided by the National Academic Infrastructure for Supercomputing in Sweden (NAISS) at PDC (DARDEL), Royal Institute of Technology, Stockholm. We have used the following Python packages to manipulate the simulation outputs and plot the results: {\tt numpy}\footnote{\url{https://numpy.org/}} \citep{harris2020array}, {\tt scipy}\footnote{\url{https://scipy.org/}} \citep{2020SciPy-NMeth}, {\tt matplotlib}\footnote{\url{https://matplotlib.org/}} \citep{Hunter2007}, and {\tt tools21cm}\footnote{\url{https://github.com/sambit-giri/tools21cm}} \citep{Giri2020}.


\section*{Data Availability}
The data underlying this article will be shared on a reasonable request to the corresponding authors.



\bibliographystyle{mnras}
\bibliography{example} 




\appendix

\section{Clustering of sources and anisotropic emission of ionising photons}
\label{sec:clustered_sources_emission}

\begin{figure}
    \centering
    \includegraphics[width=1.0\columnwidth]{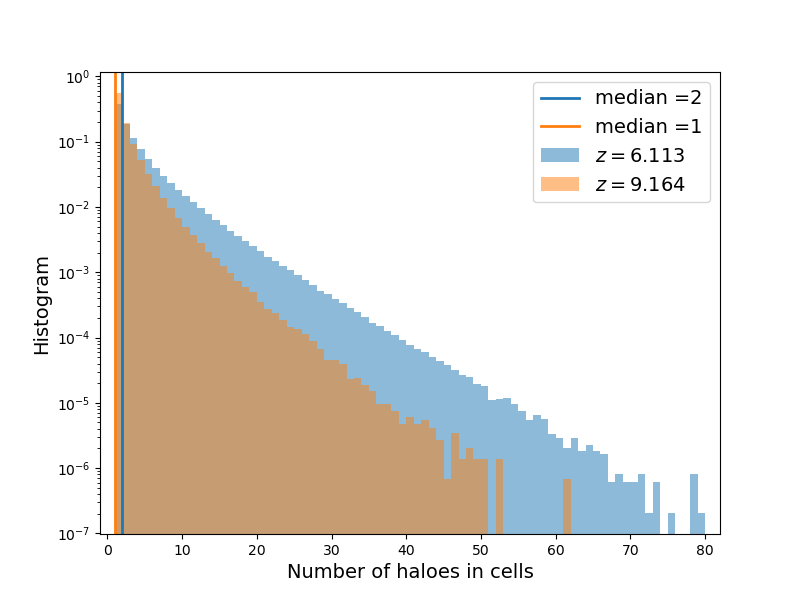}
    \caption{Normalised histogram of the number of dark matter haloes within each of the cells in our simulation volume at two redshifts marked in the legend.
    }
    \label{fig:histogram_sources}
\end{figure}

\begin{figure}
    \centering
    \includegraphics[width=1.0\columnwidth]{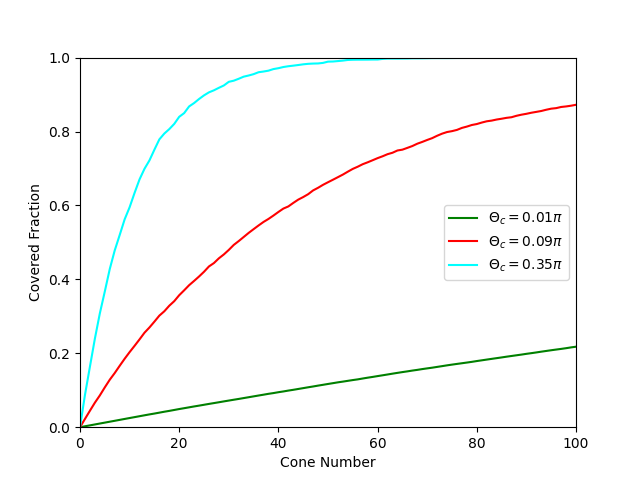}
    \caption{The average fraction of the solid angle of a sphere filled by increasing numbers of randomly-facing and overlapping cones of three different solid-angle sizes. 
    These curves show the mean value determined by repeating the entire process 50 times. The uncertainty on all these curves is less than 1 per cent.
    }
    \label{fig:fraction_overlap}
\end{figure}

The cosmological reionisation simulations used within this body of work resolve down to scales of $\sim$1 \hmpc{}. We study the impact of emission of ionising photons through narrow channels, which was seen in several high-resolution hydrodynamical simulations \citep[e.g.,][]{cen2015quantifying,paardekooper2015first,trebitsch2017fluctuating}. As we cannot resolve the small ($\sim$kpc) scale structures, we sum the contribution of ionising photons from our sources at the grid cell resolution. This method is consistent for sources with isotropic emission, however, limiting the ionising photons within a single cone and direction can underestimate the role of the clustering of the sources as well as the variance of their individual anisotropic emissions.

In order to understand the clustering of the sources within our simulation cells, in Fig.~\ref{fig:histogram_sources} we show the histogram of number of dark matter haloes per cell at $z= 9.164$ and 6.113.
These redshifts roughly correspond to global ionisation fractions of 0.1 and 1, thus bracketing most of the EoR evolution for all models studied in our main body of work (see top panels of Fig.~\ref{fig:Slices}). We observe that while most cells have few or no haloes, there are a number of cells containing many haloes, as many as 60+ at $z=9.164$ and up to 80 at $z=6.113$. Yet when comparing the median values of both distributions, these source cells account for a very small fraction of the total. 

We perform a simple test to gauge how the anisotropy of ionising photon emission by multiple sources will be imprinted at $\sim$1 \hmpc{} scales. 
We randomly filled a sphere with cones of a height the same as the radius of the sphere and the vertex at the centre of the sphere. 
With each new cone added, we kept track of the total volume of the sphere these cones were covering, adding a total of 100 cones. This was repeated 50 times and the average volume covered calculated. 
The results are visualised in Fig.~\ref{fig:fraction_overlap}, where we present the fraction of the sphere filled with cones. A value of unity corresponds to a spherical emission into the IGM due to multiple sources clustered within $\sim$1 \hmpc{} scale. 
The first two cases shown in the figure, seen in blue and red, correspond to two different assumptions for the conical emission by each source shown in Tab.~\ref{tab:models}, the \coneone{} and \conetwo{} models, respectively. We can draw useful parallels by comparing results in Fig.~\ref{fig:fraction_overlap} with the top panels of Fig.~\ref{fig:Slices} by considering how the clustering effect will impact the EoR topology. In our simulations, locations with a large / small amount of ionising sources are often interconnected (over-dense) / isolated (under-dense) regions, respectively. Note that opening angles in both the \coneone{} and \conethree{} models are identical, as the difference between the two models is the orientation of the cone (see Section~\ref{sec:method_conical}).

Comparing with the median of the distribution shown in  Fig.~\ref{fig:histogram_sources} (approximately 1 or 2 sources per cell) to the different models in Fig.~\ref{fig:fraction_overlap}, we note stark differences in the covering fraction of a factor of three when comparing the \coneone{} to the \conetwo{} case. This is roughly consistent with the different solid angles between the two models. Moreover, cells with only a few sources 
can be associated with under-dense locations of the simulations such as the bottom right of each panel of Fig.~\ref{fig:Slices}. Hence in low-density environments the role of the anisotropic emission is more pronounced, and will remain so despite the multiple separate ionising cones. 
    For values of the number of sources above 20 sources in Fig.~\ref{fig:fraction_overlap} we note that the covering fraction is close to 80 per cent for the \conetwo{} and \conethree{}{} cases, however this fraction is never reached for the  \coneone{}{} cases at the maximum number of sources. We can associate these with the largest ionised clusters in the top panels of Fig.~\ref{fig:Slices} (centre right, for example) where the number of clustered sources will be larger due to the larger over-densities. When visually compared most clusters in all cases are essentially spherical, which in the main models is a result of the large amount of bubble overlap. In addition, if we account for the decomposition of the single cone model used in the main body of the paper into multiple cones, this will boost the spherical approximation further, yet, features of the anisotropic emission might remain if the cones' opening angles are small, such as in the \coneone{} and \conethree{} cases.
    
In summary, we consider a case where each source within a cell has a separate conical emission model. We find that the opening angles of the cones will have a pronounced effect on the topology of the EoR. The effect is more pronounced in areas of the simulation with high density, however, in such areas significant overlap of ionised regions might play a more dominant role in removing the asymmetric signature of the cone (see Section~\ref{sec:results_bsd}). Conversely, the effect of the clustering is present but is less pronounced for under-dense regions which have fewer sources in general. 

\section{Resolution-Dependent Model Limitations} 
\label{sec:resolution_b}

In order to make up for the number of photons lost outside of the cone we used the boost factor $B$, the ratio of the volume of a sphere over that of a cone of a given solid angle. In these cases, $V_s(r)/V_c(r) = B$. 
This means each cell within the cone effectively receives $B$ times as many photons compared to the isotropic case. As equal amounts of photons are released on the grid, we would expect an identical ionisation history regardless of the emission model.
However, as can be seen in Fig.~\ref{fig:Reionization} ionisation is slower for narrower cones. 
This reason for this is due to the resolution of the simulation rather than recombination losses, which are in fact higher for the \sphere{} case, as seen in the ratio between the mass- and volume-averaged ionisation fraction in the bottom panel of Fig.~\ref{fig:Reionization}.

In theory, the number of cells within a sphere ($N_\mathrm{s}$) over the number of cells within a cone ($N_\mathrm{s}$) of equal radii should be $B$. 
However, this is not the case and instead the value is always slightly larger than $B$, with the difference between the two values being greater at lower resolutions and narrower cones. 
In order to confirm this we ran a Monte-Carlo-type simulation where we modelled 10,000 cones, each with a different random vector, of two different solid angles and at four separate resolutions. 
This results for this can be seen in Table~\ref{tab:mc_resolution}, where we see how the number of cells within a sphere over the number of cells within a sphere is affected by resolution.
What this means is, in order for the simulations to ionise at exactly the same rate the $B$ values should be slightly larger for narrower cones. 

However, as this is a Monte-Carlo simulation, it is important to point out that the 10,000 calculated values for $N_\mathrm{s}/N_\mathrm{c}$ have a very broad range, and are directly dependent on the vectors. 
As such, using $B = 4\pi/\Theta$ is still the best way to boost the ionising photon production efficiency in the conical models' simulations.

\begin{table}
    \centering
    \caption{The impact of resolution on the ratio between the cell within the sphere and the cells within the cone $N_\mathrm{s}/N_\mathrm{c}$.}
     \begin{tabular}{|| c c c ||} 
     \hline
     Radius & $\Theta_c$ = 0.09$\pi$ &  $\Theta_c$ = 0.35$\pi$ \\
     &(B = 44.779) & (B = 11.451) \\
     \hline
     10 & 50.079 & 11.725 \\
     20 & 45.217 & 11.466 \\
     50 & 44.794 & 11.449 \\
     100 & 44.779 &11.451

     \end{tabular}
    \label{tab:mc_resolution}
\end{table}

\section{Toy models of spherical and conical ionising photon emission} 
\label{sec:simple_case}

\begin{figure}
    \centering
    \includegraphics[width=\columnwidth]{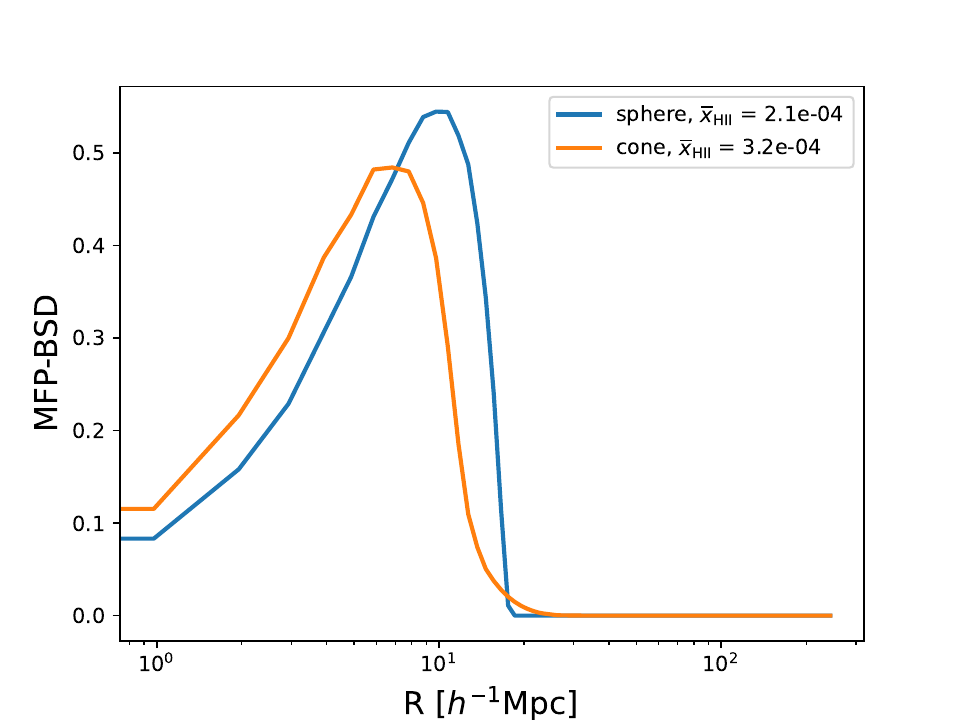}
    \includegraphics[width=\columnwidth]{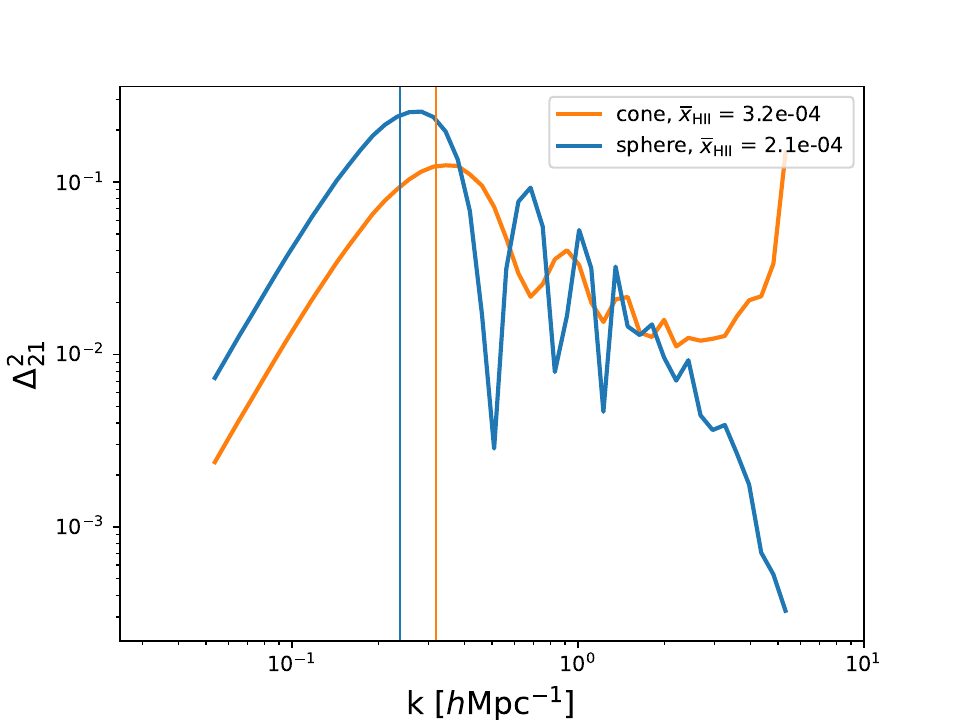}
    \includegraphics[width=\columnwidth]{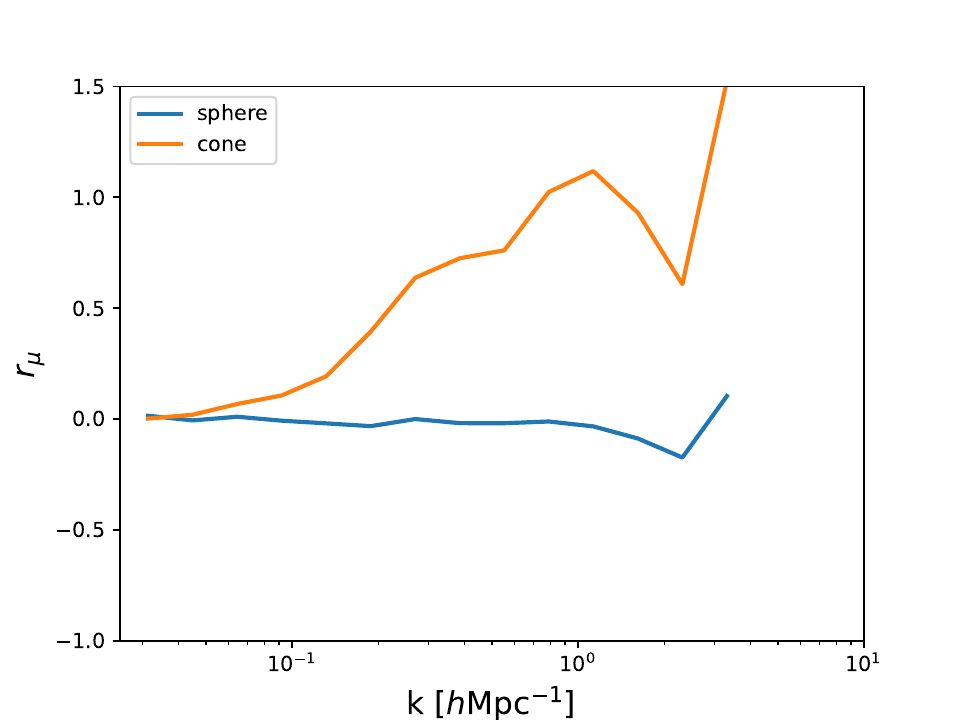}
    
    \caption{Comparison of the MFP-BSDs (top panel), the power spectra (middle panel) and the anisotropy ratio (bottom panel) of the ionisation field, generated for the cases with a single spherical (blue) and single conical (orange) ionised region, which are illustrated in Fig.~\ref{fig:One_Cone}. 
    }
    \label{fig:Simple_cases}
\end{figure}

In order to better understand how the anisotropy of ionising photon emission impacts our analysis, we examine the toy model shown in Fig.~\ref{fig:One_Cone}. In this case, only two ionising sources are introduced in a uniform IGM, with a hydrogen number density equal to the mean value at the given redshift. The source visible on the bottom left of the figure follows the fiducial spherical emission model, and the source on the top left follows the anisotropic prescription described in Sec.~\ref{sec:method}. As discussed in Sec.~\ref{sec:results_mfp_bsd}, ionised regions overlap even during the early stages of reionisation, creating complex topologies. This complicates the task of discerning the impact of the chosen emission model. Below we analyse the two single source cases from Fig.~\ref{fig:One_Cone} separately, considering the MFP-BSD (Sec.~\ref{sec:results_mfp_bsd}), the 21-cm power spectrum (Sec.~\ref{sec:results_ps}) and the anisotropy ratio (Sec.~\ref{sec:results_rmu}).

Fig.~\ref{fig:Simple_cases} shows the MFP-BSD calculated according to the algorithm from \citet{Mesinger2007} for single anisotropic and isotropic sources. The global ionisation fraction for the single cone case is slightly larger than that of the sphere and this can be attributed to the fewer partially ionised cells around the cone, as shown in Fig.~\ref{fig:One_Cone}, which shift the distribution. These results confirm that in the single cone case (compared to the spherical case), the topology of the cone results in more lines of sight smaller than the peak. This was also seen in early EoR results in Fig.~\ref{fig:BSDs}. Moreover, the elongated shape of the cone in Fig.~\ref{fig:One_Cone} results in fewer longer lines-of-sight in the distribution. This leads to lower numbers for $R>10$ \hmpc{} in the top panel of Fig.~\ref{fig:Simple_cases}. In particular, the peak of the MFP-BSD distribution for the single cone model $R_{\mathrm{peak,c}} = 7.3 \pm 3.6\, h^{-1}$Mpc is approximately 20-25 per cent lower than that of the single sphere $R_{\mathrm{peak,s}} = 9.3 \pm 3.8\, h^{-1}$Mpc.
While this is easily confirmed by the visual inspection of the radius of the sphere ($R_{\mathrm{sphere}} = 8.4\, h^{-1}$Mpc) in Fig.~\ref{fig:One_Cone}, we note that the radius of the hemispherical cap of the cone ($R_{\mathrm{sphere}} = 6.3\, h^{-1}$Mpc) is the main feature picked up by the MFP-BSD analysis.

In the middle panel of Fig.~\ref{fig:Simple_cases}, we present the dimensionless spherically-averaged power spectrum of the 21-cm signal field for each simplified case, where the vertical lines show the imprint of the bubble radius on the power spectrum $k_{\mathrm{c}} \approx2/R$ \citep[see][for a discussion]{georgiev2022large}.
Comparing to the top left panel of Fig.~\ref{fig:PS_clean} (early on in the EoR) we re-affirm the findings outlined in Sec.~\ref{sec:21cm_ps}. We again see that the amplitude of the 21-cm power spectrum is higher for the spherical case since it can produce an ionised bubble with a radius larger than that of the hemispherical cap of the cone. Conversely, it is the single cone that has a higher amplitude for small physical scales (high-$k$ scales) due to the elongated body of the cone. 

These geometric properties are now more clearly visible in the shape of each power spectrum as there are no overlapping regions, such as in Fig.~\ref{fig:PS_clean}. Notably, the peak of the power spectrum is also shifted by 20-25 per cent, consistent with the MFP-BSD. Moreover, the power spectrum of the single sphere has a distinct parabolic shape defined by the peak of the ionised bubble, while that of the single cone tends to have more power for scales larger than $k_{\mathrm{c}}$. This is conceptually in line with our findings, however, the simplistic model is incapable of capturing the higher order contributions of the 21-cm power spectrum, which arise from the coupling of the density and ionisation fields \citep[see eq.~2 of][]{Lidz2007}.

Lastly, we examine the anisotropy ratio from Sec.~\ref{sec:results_rmu} (also see Eq.~\ref{eq:anisotropy_ratio}) in the bottom panel of Fig.~\ref{fig:Simple_cases}.
We note a clear anisotropy feature in the case of the single cone, which increases with the $k$-scale. 
As discussed in the case of the power spectrum, the small physical scales (large-$k$ scales) are more sensitive to the body of the cone, and hence its anisotropic shape. However, when compared to our simulation outputs in Fig.~\ref{fig:rmu_clean}, the anisotropy induced by the anisotropic emissions is significantly suppressed.
This is best understood when comparing the ionisation slices of the \sphere{} and \coneone{} cases
in the upper panel of Fig.~\ref{fig:Slices}. Even early in the EoR, the number of overlapping ionised structures in the \coneone{} case significantly reduce the anisotropic emission of its ionising sources. Therefore, despite the individual anisotropy introduced by the conical emission of each source, the cumulative effect on the anisotropy ratio is isotropic.

\section{Threshold effects in the Friends-of-Friends Bubble Size Distributions} 
\label{sec:ionisation_limit}

\begin{figure*}
    \centering
    \includegraphics[width=1.0\linewidth]{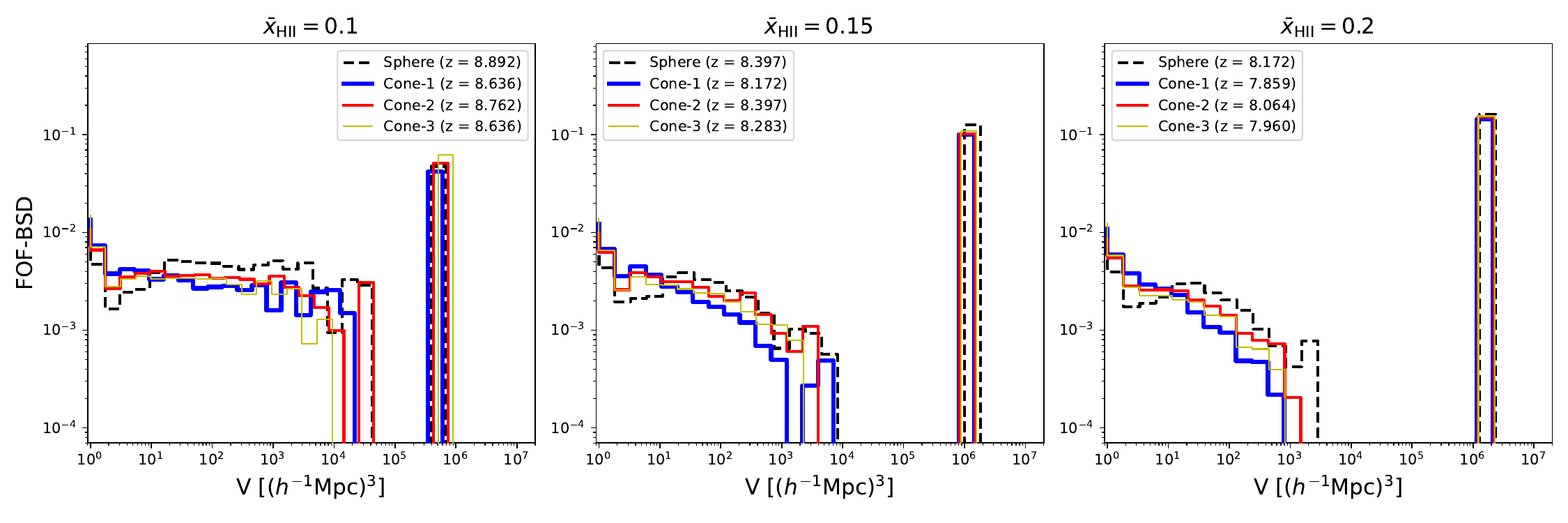}
    \includegraphics[width=1.0\linewidth]{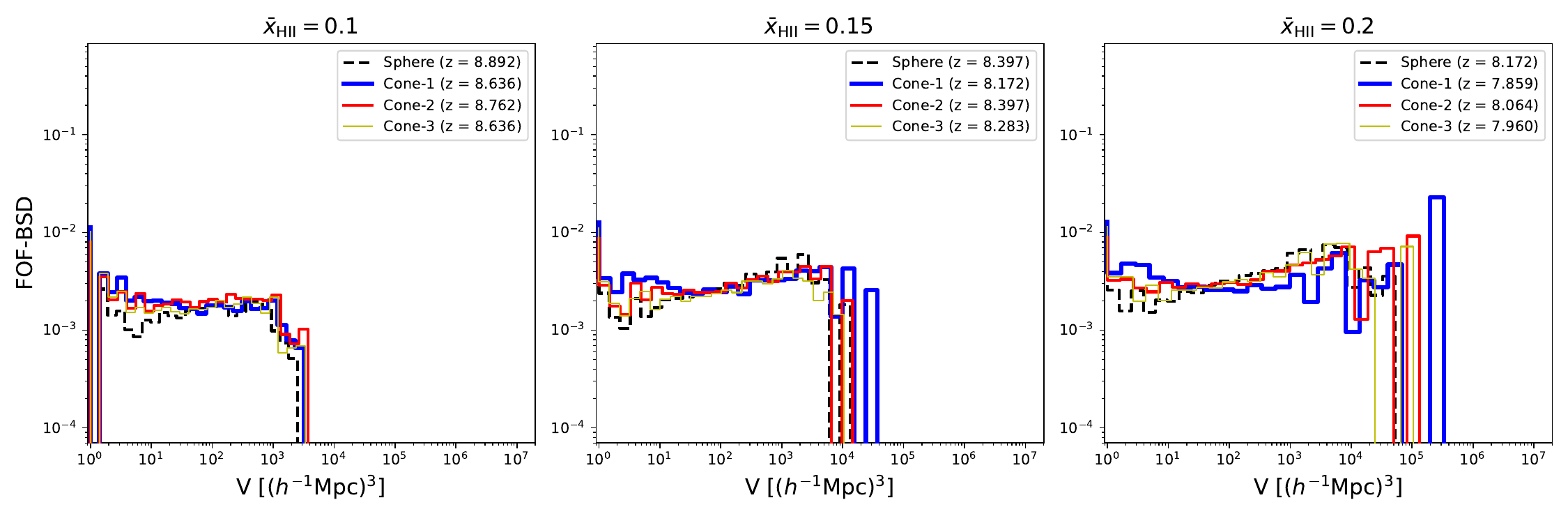}    
    \caption{Comparison of the FOF-BSDs at ionisation thresholds of  $x_\mathrm{HII}\ge 0.1$ and $x_\mathrm{HII}\ge 0.9$ in the top and bottom panels, receptively. 
    }
    \label{fig:comp_10_90}
\end{figure*}

In Sec.~\ref{sec:results_fof_bsd} we used the threshold of ($x_\mathrm{HII}\ge 0.5$) to determine if a cell was considered ionised or not when measuring the FoF-BSD. 
However, something of note here was that in the conical cases the radiation would always be more concentrated. A single cell would be receiving $B$ times as much radiation as in the \sphere{} case, while the number of cells receiving this increased radiation would be fewer in the conical cases since a sphere has a greater surface area than that of the curved end of a cone of equal volume. 
This would suggest that there is a significant difference between fully and partially ionised cells in the spherical and conical cases.  
The effect of this can be seen Fig.~\ref{fig:comp_10_90} where recreate the FOF-BSDs in the top panels of Fig.~\ref{fig:BSDs} for ionisation thresholds of $x_\mathrm{HII}\ge 0.1$ and $x_\mathrm{HII}\ge 0.9$ for the top and bottom panels, respectively. 
Naturally, the ionised regions present for the threshold of 0.9 are smaller and fewer than those for a threshold of 0.1. However, one thing of note here is how the bubbles sizes shift relative to the case in question. 

In the upper panels where $x_\mathrm{HII}\ge 0.1$, we seen that the mid-sized bubbles (V > 10$^1$ (h$^{-1}\mathrm{Mpc})^{3}$) are fewest in number in the \coneone{}, this being the case where the radiation is most concentrated and, unlike in the \conethree{} case, does not change direction over time which results in fewer cells being only partially ionised. 
We can also see that the mid-sized ionised regions are the quite significantly larger in the \sphere{} case, and indeed this case also results in the largest, single-volume formed by percolation. 
The top-right panel nicely demonstrates this, with the \sphere{} case showing the largest bubbles, then the less-concentrated \conetwo{} case, followed by the less the concentrated, but varying over time, \conethree{} case and then the \coneone{} case with concentrated radiation and static vectors. 

Looking at the bottom panels, where a cell must be almost fully ionised to be considered, we see some noticeable differences. Firstly, percolation has not yet occurred and the largest bubble is formed in the \coneone{} case, almost a full magnitude larger than in the \sphere{} simulation.
Between these two extreme cases, we see the other two cases' largest bubble sizes. 
These are the also larger, but not to the same level. 

In conclusion, we see that there is a significant difference in the number and spacial distribution for the FoF-BSD when comparing fully and partially ionised cells.

\label{lastpage}

\end{document}